%% file: 4223.tex
\documentclass[a4paper]{aa}
\usepackage{graphicx,txfonts,aalongtable}

\newcommand{\fo}{\ensuremath{f^\mathrm{o}}}
\newcommand{\fe}{\ensuremath{f^\mathrm{e}}}
\newcommand{\Den}{\ensuremath{f^\mathrm{o} + f^\mathrm{e}}}
\newcommand{\Bz}{\ensuremath{\langle B_z\rangle}}
\newcommand{\te}{\ensuremath{T_{\mathrm{eff}}}}

\newcommand{\cz}{\ensuremath{C_z}}
\newcommand{\kms}{\ensuremath{\mathrm{km\,s}^{-1}}}
\newcommand{\vsini}{\ensuremath{v_\mathrm{e} \sin i}}

\providecommand{\phip}{\ensuremath{\phi^{+}}}
\providecommand{\phim}{\ensuremath{\phi^{-}}}
\newcommand{\Mean}[1]{\ensuremath{\left\langle{#1}\right\rangle}}
\newcommand{\Var}[1]{\ensuremath{\sigma^2[{#1}]}}
\newcommand{\Covar}[2]{\ensuremath{\sigma[ {#1},{#2}]}}

\begin{document}
\title{Searching for links between magnetic fields and stellar evolution}
\subtitle{I. A survey of magnetic fields in open cluster A- and B-type stars with FORS1
\thanks{Based on observations made with ESO Telescopes at the Paranal Observatory under programme ID 
068.D-0403,
070.D-0352,
272.D-5026,
073.D-0498,
and
074.D-0488}}
       \author{
        S.~Bagnulo      \inst{1}
       \and
        J.D.~Landstreet \inst{2}
       \and
        E.~Mason        \inst{1}
       \and
        V.~Andretta     \inst{3}
       \and 
        J.~Silaj        \inst{2}
       \and
        G.A.~Wade       \inst{4}
        }
\institute{European Southern Observatory,
           Casilla 19001,
           Santiago 19, Chile.\
           \email{sbagnulo@eso.org, emason@eso.org}
           \and
           Physics \& Astronomy Department,
           The University of Western Ontario,
           London, Ontario, Canada N6A 3K7. \\
           \email{jlandstr@astro.uwo.ca, jsilaj@uwo.ca}
           \and
           INAF - Osservatorio Astronomico di Capodimonte, 
           salita Moiariello 16, 80131 Napoli, Italy.
           \email{andretta@na.astro.it}
           \and
           Department of Physics,
           Royal Military College of Canada,
           P.O. Box 17000, Station `Forces'
           Kingston, Ontario, Canada K7K 7B4.\\
           \email{Gregg.Wade@rmc.ca}
           }
\date{Received: 19 September 2005 / Accepted: 12 January 2006}

\abstract
{ 
About 5\,\% of upper main sequence stars are permeated by a strong
magnetic field, the origin of which is still matter of debate.
}
{
With this work we provide observational material to study
how magnetic fields change with the evolution of stars on the main
sequence, and to constrain theory explaining the presence
of magnetic fields in A and B-type stars.
}
{
Using FORS1 in spectropolarimetric mode at the ESO VLT, we have carried out
a survey of magnetic fields in early-type stars belonging to open
clusters and associations of various ages. 
}
{
We have measured the magnetic field of 235 early-type stars with a
typical uncertainty of $\sim 100$\,G.  In our sample, 97 stars are Ap
or Bp stars. For these targets, the median error bar of our field
measurements was $\sim 80$\,G. A field has been detected in about 41
of these stars, 37 of which were not previously known as magnetic
stars.  For the 138 normal A and B-type stars, the median error
bar was 136\,G, and no field was detected in any of them.
}
{}
\keywords{Stars: magnetic fields --
          Stars: chemically peculiar -- 
          Stars: evolution --
          Polarization -- 
          Techniques: polarimetric  }
\titlerunning{A survey of magnetic fields in open cluster A- and B-type stars with FORS1}
\authorrunning{S.~Bagnulo, J.D.~Landstreet, E.~Mason, et al.}

\maketitle


\section{Introduction}\label{Sect_Introduction}
The presence of strong ($\sim 1$\,kG) magnetic fields in some of the
A- and B-type stars of the upper main sequence has been known for more
than 50 years (Babcock \cite{Bab47}). These fields are almost
invariably associated with a suite of other unusual characteristics,
which include (a) specific angular momentum of the order of 10\,\% or
less of the typical value for normal main sequence stars of similar
mass; (b) quite anomalous atmospheric chemical composition, which is to
first approximation a function of effective temperature (and thus is
apparently an atmospheric rather than a global feature); and (c)
variation of the spectrum, light and magnetic field with the star's
rotational period, that clearly indicate the presence of quite
significant abundance inhomogeneities and of a magnetic field not
symmetric about the rotation axis.

Although considerable progress has been made in understanding the
physical processes at work in these stars, many important problems
remain unsolved. Among these are two major questions.  First, although
there is strong evidence (e.g., the stability of the observed fields,
the lack of symptoms of Sun-like activity, and the lack of any
important correlation of observed field strength with rotational
angular velocity) that the observed fields are fossil fields, it is
not yet clear how these fields evolve during the main sequence
phase. Secondly, although it is believed that the basic mechanism
leading to both chemical anomalies and to atmospheric inhomogeneities
is the competition between gravitational settling, radiative
levitation, and various hydrodynamic processes, the interplay of these
processes is still very poorly understood.

In this situation, it is helpful to look to observations to guide
physical theory. One kind of information about the magnetic Ap and Bp
stars (hereafter referred to as magnetic Ap stars) that has been
almost entirely lacking is the age of observed stars. Good age
information would be very useful for discerning systematic
evolutionary changes in field strength, chemical composition, rotation
rates, etc.  The general lack of useful age information about magnetic
Ap stars occurs because almost all of the bright magnetic Ap stars are
field stars. Even with the accurate parallaxes now available from the
Hipparcos mission, the uncertainties in luminosity and effective
temperature of Ap stars are large enough that placing them in the HR
diagram only suffices to determine very roughly their stage of
evolution (see Sec 2.1 below).

The obvious way to obtain improved ages for magnetic Ap stars is to
observe such stars in open clusters. Until recently such a study has
not been possible because cluster Ap stars are mostly fainter than
$V \approx 6$ or 7, beyond the limit of accurate magnetic field
measurements with available instruments. This situation has changed
due to the development of a new generation of highly efficient
spectropolarimeters and observing strategies, and their availability
on large telescopes. In particular, the FORS1 spectropolarimeter on
one of the ESO 8\,m VLT telescopes has been shown to be a powerful
tool for measuring fields in very faint Ap stars. It has recently been
used to detect a field in a star of $V = 12.88$, the faintest magnetic
Ap in which a field has ever been detected (Bagnulo et al.\
\cite{Bagetal04}).

Another very important development has been the substantial increase
in the number of probable magnetic Aps identified in clusters,
particularly by the systematic surveys of Maitzen and his
collaborators. Furthermore, the availability of very accurate proper
motions for a very large number of stars from the Hipparcos mission
and the Tycho-2 catalogue has greatly facilitated the correct
separation of cluster members from foreground and background stars.

The time is now clearly ripe for studying magnetic Ap stars in
clusters to obtain for the first time a reasonably large sample of
magnetic Ap stars of known absolute and evolutionary ages. We have
started to carry out such a survey, using the FORS1
spectropolarimeter. The first stage of this survey is reported in this
paper. Section~\ref{Sect_Scope} discusses the rationale and scope of
this survey, in particular what the advantages are of studying open
cluster stars compared to studying field stars, why FORS1 at the VLT
is an ideal instrument for this survey, and how individual targets
have been selected.  Sections~\ref{Sect_Basic} --
\ref{Sect_Balmer_Metal} describe how the magnetic field can be
determined from observations of polarized spectra in terms of basic
physics, observing strategy, data reduction, and \Bz\
determination. These sections contain, for other FORS1 users, a
detailed discussion of the optimized techniques we have developed for
field measurement, and may be skipped by readers interested mainly in
the observational results.  In Sect.~\ref{Sect_Observations} and
\ref{Sect_Discussion} we present and discuss the observations obtained
during this survey. Conclusions are presented in
Sect.~\ref{Sect_Conclusions}.

\section{The scope of this survey}\label{Sect_Scope}

\subsection{Why open cluster stars?}\label{Sect_Why_Open}
The final goal of this series of papers will be to obtain
ages, masses, and magnetic field strengths for a substantial
number of Ap stars, and study whether there are evolutionary changes
in the magnetic field strength. The magnetic (and sometimes the
abundance) properties of a large sample of Ap stars in the field
around the Sun are already known (e.g., Mathys \cite{Mathys04}, Cowley
\& Bord \cite{CowBor04}). In principle, our evolutionary study
could be based on these data of field stars. To justify the need of
the present survey of magnetic cluster stars we first consider the
extent to which ages and masses may be derived for the stars in the
field.

Temperatures for many of field stars may be estimated using
available \textit{UBV}, \textit{uvby}, or Geneva photometry, together
with calibrations provided by St\c{e}pie\'{n} \& Dominiczak
(\cite{SteDom89}), Hauck \& North (\cite{HauNor93}), and Napiwotzki,
Sch\"{o}nberner, \& Wenske (\cite{Napetal93}). Luminosities require
distances and bolometric corrections. Distances to many nearby Ap
stars have recently been accurately determined by the Hipparcos
project (see Gomez et al. \cite{Gometal98}); for the Ap stars nearer
than about 100 pc, the relative distance errors are less than about
10\%. The bolometric corrections required have been discussed by Lanz
(\cite{Lanz84}). Thus, it is now possible to place a large number of
Ap stars on the HR diagram, and, using the evolution tracks for stars
of various masses, to estimate both the mass and the age of individual
stars. This exercise has been carried out by Gomez et
al. (\cite{Gometal98}), by P\"{o}hnl et al.\ (\cite{Poeetal05}),
and by Kochukhov \& Bagnulo (\cite{KocBag06}) for a very
large sample of nearby Ap stars, and by Hubrig, North, \& Mathys
(\cite{Hubetal00}) for a special sample of particularly slowly
rotating magnetic Ap stars. In fact, as discussed below, this method
is limited by large uncertainties.

Effective temperatures of magnetic Ap stars are still somewhat
uncertain, more so than for normal main sequence stars because of the
peculiar energy distributions of Ap stars (see St\c{e}pie\'{n} and
Dominiczak \cite{SteDom89}). A recent tabulation by Sokolov
(\cite{Sokolov98}) of effective temperatures for nearly 70 Ap stars,
with comparisons to earlier determinations, suggests that the
uncertainty in \te\ typically about 5\,\%, or $\pm 0.02$
dex.

\begin{figure*}
\resizebox{12cm}{!}{
\includegraphics*[angle=90,width=9cm]{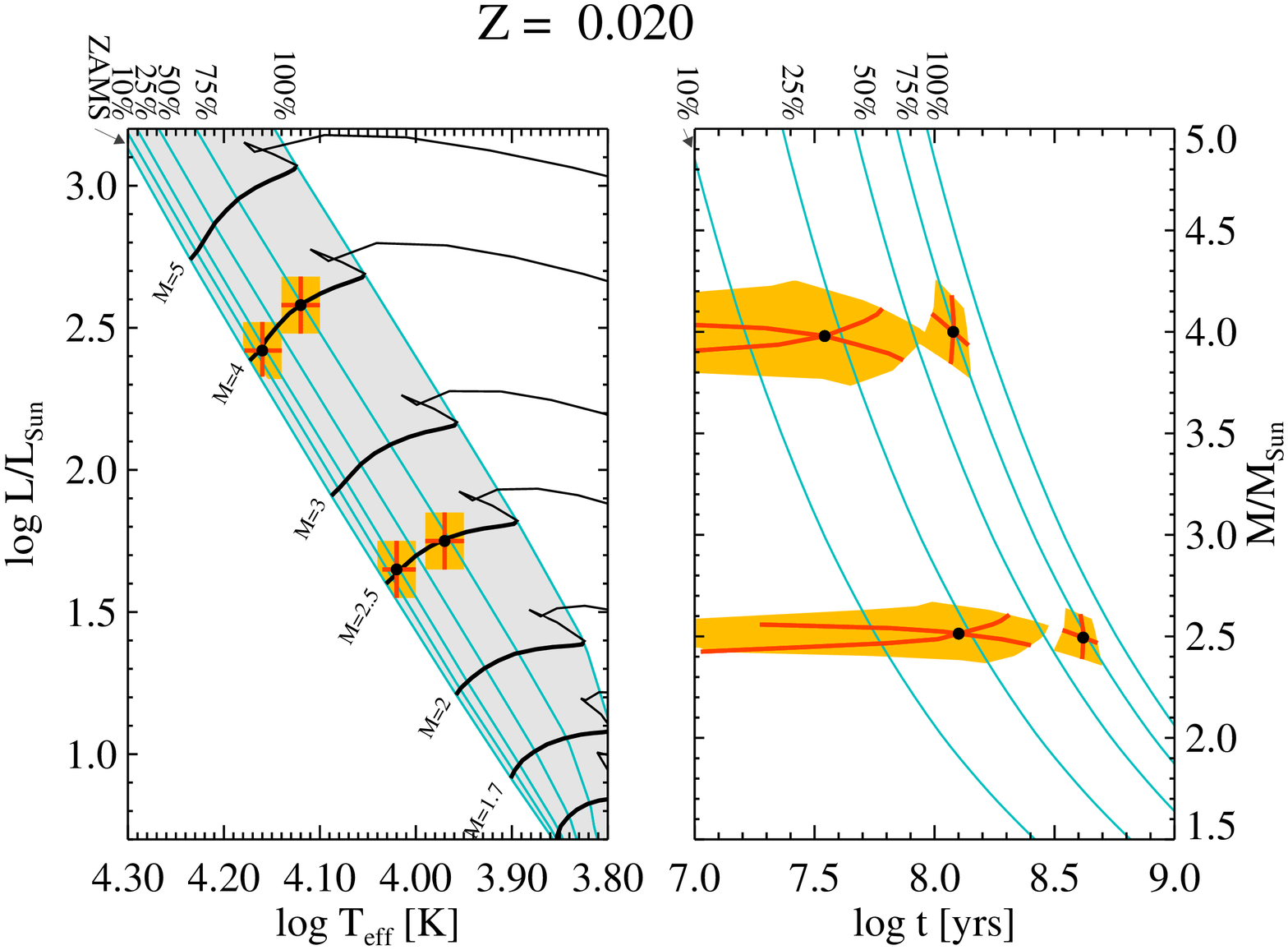}}\\
\resizebox{12cm}{!}{
\includegraphics*[angle=90,width=9cm]{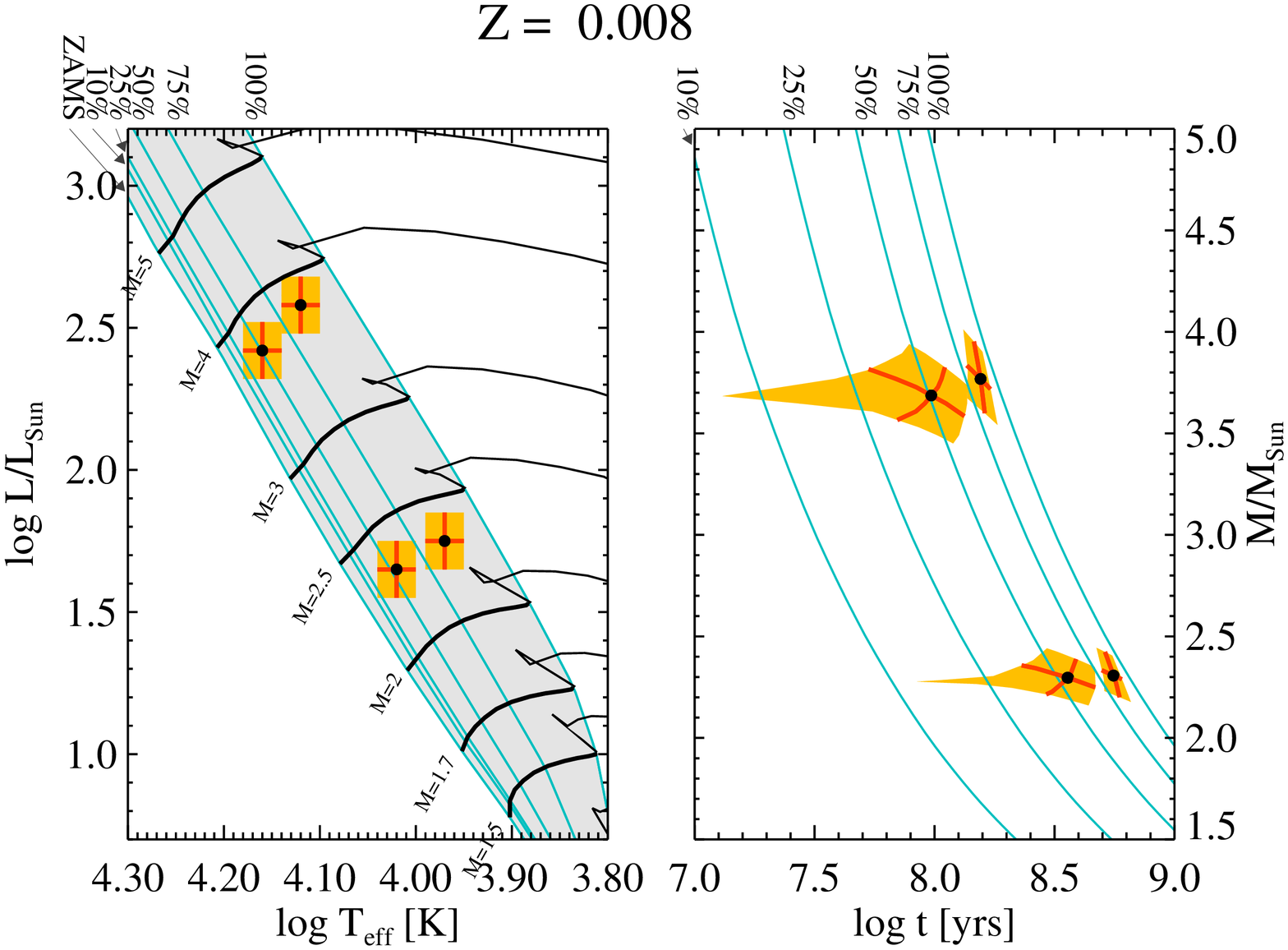}}
\hfill
\parbox[b]{55mm}{
\caption{\label{Fig_mass-age} Top panels: the position of a star in
the HR diagram, and the star's position transformed into a diagram of
age as a function of stellar mass, assuming that we know effective
temperature $\pm 5$\% and luminosity $\pm 0.1$ dex. The transformation
uses standard evolution tracks for $Z = 0.02$ (Schaller et
al. (\cite{Schetal92}); several fractional ages (fraction of main
sequence life completed) are labelled.  Bottom panels: the same
transformation as in the top panels, but using tracks of metallicity
$Z=0.008$ (Schaerer et al.\ \cite{Schetal93})}}
\end{figure*}

For Hipparcos parallaxes, the distance uncertainty at 100 pc is about
8\,\% (or about 0.035\,dex), and increases with the distance
itself. There is also an uncertainty in the appropriate bolometric
correction (BC) to apply. First of all, there is an uncertainty of
$\sim 0.1$\,mag in the estimate of the BC of a normal A-type star (due
in part to the uncertainty in \te). Then, for Ap stars, one has to
apply a {\it correction} to the normal BC. According to Lanz
(\cite{Lanz84}) this correction is uncertain by $\sim
0.2$\,mag. Finally, there is the uncertainty as to whether one should
apply the Lutz-Kelker correction, about 0.1\,mag at 100 pc. Taking
these effects together, for nearby field stars, the uncertainty in
$M_{\mathrm{bol}}$ is $\sim 0.3$\,mag, i.e., $\sim 0.1$\,dex in
$\log(L/L_\odot)$.  These uncertainties, once converted in the ($\log
t$,$M/M_\odot$) plane, correspond to age uncertainties that may be as
large as half the main sequence life.

Four examples of this kind of inference are shown in the upper panel
of Fig.~\ref{Fig_mass-age}, for stars of about $2.5\,M_\odot$ and
$4.0\,M_\odot$, each considered for an age near the beginning of its
main sequence lifetime and near the end. Comparison is made with the
tracks of Schaller et al.\ (\cite{Schetal92}) for $Z = 0.02$. The star
positions in the HR diagram, with their uncertainties (shaded boxes), are
shown in the upper left panel. The deduced positions in the age-mass
diagram are shown in the upper right panel. It is clear that the
error boxes for observational characteristics of Ap stars translate
into ages which are sufficiently uncertain that {\em one 
cannot resolve at least the first half of the main sequence
lifetime}. The situation is somewhat better near the end of the main
sequence life, where the isochrones are farther apart, but typically
an age uncertainty of the order of $\pm 25$\,\% of the total main sequence
lifetime is to be expected.

A further uncertainty in the transformation from the ($\log\te$,
$\log(L/L_\odot)$) to the ($\log t$, $M/M_\odot$) plane comes from the
fact that we do not know accurately the bulk chemical composition of
any field Ap star, and thus we do not know what chemical composition
should be assumed in the theoretical tracks used for the
comparison. To estimate the size of this effect, we consider the range
of abundances ([Fe/H]) present in open clusters young enough to still
have Ap stars (say $\log t < 9$). Searching the WEBDA database
(e.g. Mermilliod \& Paunzen \cite{MerPau03}), we found that this range
is of the order of 0.4\,dex. This result is confirmed by examination
of the younger clusters in the catalogue of Chen, Hou, \& Wang
(\cite{Cheetal03}). This range of abundances may represent a
reasonable estimate of the range of values of [Fe/H] present in nearby
field stars.

A change in [Fe/H] of 0.4\,dex 
corresponds to a change in metal abundance of about a factor of
2.5. This is just the difference in abundance between the evolution
calculations with $Z = 0.008$ and $Z = 0.02$ provided by the Geneva
group (Schaller et al.\ \cite{Schetal92}; Schaerer et al.\
\cite{Schetal93}). Repeating the transformation from the ($\log\te$,
$\log(L/L_\odot)$) to the ($\log t$, $M/M_\odot$) plane for $Z =
0.008$, we find the results shown in the lower panels of
Fig.~\ref{Fig_mass-age}. (We have used the same values of \te\ and
$\log(L/L_\odot)$ as in the upper panels.) The effect of the
uncertainty in bulk chemical composition is to add roughly another
$\pm 0.25$ uncertainty to the deduced fraction of the main sequence
life completed by a particular star. Overall, we see that knowledge of
\te\ and $\log(L/L_\odot)$ provides us with mass estimates accurate to
about $\pm 10$\%, but provides only poor age resolution, especially in the
first half of the main sequence life (see also Fig.~4 of 
Kochukhov \& Bagnulo \cite{KocBag06}).

The situation is substantially improved if the star is a member of a
cluster or association. The absolute age of the star is then known
with essentially the precision of the cluster age. The uncertainty of
this number varies from cluster to cluster, mainly because of the
difficulty of deciding exactly where to place the cluster turnoff. 
Typically the uncertainty in
$\log t$ is of the order of $\pm 0.2$ dex, about $\pm 50$\% of the
absolute age (see for example Castellani et al. \cite{ccs92}),
although more accurate ages are sometimes reported (e.g. Carrier et
al. \cite{cbr99}). This represents a very considerable improvement in
absolute age, especially for a star in the early part of its main
sequence life. If we know only position in the HR diagram, a star with
an actual age of $10^7$\,yr and a main sequence lifetime of $10^9$\,yr
would have an age uncertainty of the order of $3\, 10^8$\,yr, while
the same star in a cluster would have an age uncertainty of roughly
$3\,10^6$\,yr. Only for a star in about the last third of its main
sequence life is the age uncertainty not substantially improved by
knowing that it is in a cluster. Another advantage of studying cluster
stars is that one may determine the appropriate cluster bulk
metallicity (and hence decide what evolution tracks to use for
comparison) by studying the lower main sequence stars.

The masses of cluster stars can be determined with similar methods and
similar (or better) accuracy than for local field stars. The values of
\te\ are known with about the same accuracy as for local field stars,
and there are the same uncertainties for the bolometric correction.
Hipparcos parallaxes are normally not available, but the value of
luminosity is obtained from the observed $V$ magnitude together with
the cluster's apparent distance modulus $(V - M_V)$. Recent
determinations of distance modulus by main sequence fitting appear to
achieve an accuracy of about $\pm 0.2$\,mag (e.g., Robichon et
al. \cite{Robetal99}). This accuracy is generally obtained out to
distances well beyond those few clusters for which accurate ($\pm 25$\%)
parallaxes are available. Furthermore, if we know the cluster
metallicity, we can decide which evolution tracks to use. The
precision of the mass determination of cluster stars is about $\pm
10$\% if the bulk composition is not known, and somewhat better if it
is.

Our conclusion is that an age derived using only the position in the
HR diagram of an \textit{individual} Ap star in the field is at
present sufficiently uncertain to be of little value except for stars
near the end of their main sequence lives (although important
conclusions can still be drawn from statistical considerations: see
P\"{o}hnl et al.\ (\cite{Poeetal05}) and Kochukhov \& Bagnulo
\cite{KocBag06}).  In contrast, if the star is a member of a cluster
of known age, it is possible to determine accurately {\em both} the
mass and age (or fraction of main sequence life elapsed). Therefore,
the study of Ap stars that are cluster members is of great value in
understanding the temporal evolution of rotation, magnetic fields, and
atmospheric chemistry in all magnetic Ap stars.

At present only a few cluster Ap star candidates are known to be
magnetic. The total number of clusters for which magnetic
observations had been published is eight, with a total of 13 stars
surveyed. In addition, extensive surveys had been carried in for the
Ori OB1 (Borra \cite{Borra81}) and Sco OB2 associations (Thompson et
al.  \cite{Thoetal87}). Accordingly, we have decided to carry out a
survey of magnetic stars in open clusters to provide age information
for a substantial sample of Ap stars.


\subsection{Why FORS1 at the VLT}\label{Sect_Low_Resolution}
To decide on a suitable observing strategy, we must determine
what field strength we wish to detect. Among Ap stars in the local
field, it appears that the median root-mean-square observed
line-of-sight field strength \Bz\ is of the order of 300\,G
(Bohlender \& Landstreet \cite{BohLan90}). We assume that this will
also be typical of cluster Ap stars. In order to detect such a small
value, it is important for the survey to achieve measurement
uncertainties of the order of 100\,G or less, as far as possible
independent of the star's effective temperature and \vsini. 

In principle, we may carry out a survey by searching either for
visible Zeeman splitting of spectral lines in a simple intensity
spectrum, or by searching for the circular polarization signature of a
global field of simple structure. Although Mathys and collaborators
(e.g., Mathys et al.\ \cite{Matetal97}) have shown that Zeeman
splitting may be detected in some tens of field stars, this requires
quite special circumstances (i.e., \vsini\ at most a few \kms, and
field strength at least 2~kG) that are not met in most of the known
magnetic Ap stars. Polarization measurement is generally a far more
sensitive and broadly-applicable method of field detection than
observation of Zeeman splitting.

Two main methods of measurement are currently in use for detecting the
circular polarization produced by a non-zero value of \Bz. One method
exploits the Zeeman polarization in metal lines (e.g., Babcock
\cite{Bab58}; Preston \& St\c{e}pie\'{n} \cite{PreSte68}; Mathys \&
Hubrig \cite{MatHub97}; Wade et al.\ \cite{Wadetal00}; Elkin,
Kudryavtsev, \& Romanyuk \cite{Elketal03}). Field measurement by this
method usually relies on circular spectropolarimetry with a resolving
power $R \geq 3\, 10^4$. If the star is quite bright, with a \vsini\
value smaller than, say, 15 or 20 \kms\ and a rich spectrum, field
errors $\sigma_B$ as small as a few G can be achieved (Shorlin et al.\
\cite{Shoetal02}). On the other hand, the measurement uncertainty
depends strongly on spectral type (which determines the number of
usable lines and their intrinsic depths) and on \vsini.  In
observations collecting similar total numbers of photons, the standard
error of field measurement can vary by a factor of order $10^2$.

A second method employs the Zeeman polarization in 
the wings of the Balmer lines. These lines may be
observed with quite low resolving power ($R \sim 10^3$), using either
interference filters (e.g., Borra \& Landstreet \cite{BorLan80}) or a
low-dispersion spectrograph (Bagnulo et al.\ \cite{Bagetal02}). Since
the Balmer lines are intrinsically broad, the polarization
signal may be a factor of 10 smaller than in the metal lines, and the
best achievable standard errors are of order 30 -- 50\,G (Landstreet
\cite{Lan82}). However, since the Balmer lines are always quite deep,
and do not vary much in strength among A and B stars, and since the
overall profile at $R \approx 10^3$ is hardly affected by rotation,
this method can provide standard errors that are fairly uniform simply
by surveying a sample of stars to a specified signal-to-noise ratio.

The (spectro-)polarimeters now in use make both kinds of field
measurement. However, limitations imposed by low overall efficiency
and (usually) modest telescope aperture have limited most of the \Bz\
field measurements to stars not much fainter than $V = 6$. This
magnitude limit has effectively prevented any serious survey of
magnetic fields in cluster Ap stars, as only a handful of clusters and
associations have a significant number of Ap stars brighter than $V =
6$. With a limiting magnetic around $V = 6$, one is restricted to
clusters and associations not much more than 100 pc away, since the
absolute magnitude of an Ap star is typically in the range of $M_V
\sim -1$ to +1. Only half a dozen clusters are this near.

The development of the FORS1 spectropolarimeter for the ESO Very Large
Telescope (VLT) has changed the situation dramatically. FORS1
incorporates a multi-object low dispersion spectrograph with
polarizing optics. In polarimetric mode, spectra with $R \sim 10^3$
may be obtained for up to nine objects simultaneously in a field $7'$
square.  Bagnulo et al. (\cite{Bagetal02}) have shown that FORS1 can
be used very effectively as a Balmer-line polarimeter for field
measurement.  Since this instrument is mounted on an 8\,m telescope,
and has a very high throughput, the limiting magnitude is much fainter
than with earlier instruments. Bagnulo et al.\ (\cite{Bagetal04}) have
used FORS1 to detect a field in a star of $V \approx 13$ in 2.8 hours
of observation, and showed that it is possible to reach a precision of
$\sigma_B \sim 10^2$ G in about 1 hour of integration at $V =
10$. With a limiting magnitude of 10 or even fainter, we can survey
clusters and associations out to several hundred pc, and the number of
clusters that can be studied rises to roughly 100.  In addition
Bagnulo et al.\ (\cite{Bagetal02}) have shown that the low resolution
does not prevent the use of metal lines to measure the magnetic field
in stars with rich metallic spectra. This fact may be used to increase
the sensitivity of field measurements in some stars. A third advantage
of FORS1 is that in clusters, it is often possible to observe several
stars simultaneously. We have used this capability to observe multiple
Ap candidate stars, but also non-Ap cluster members in the hopes of
making serendipitous field detections.

This survey is biased toward stars with longitudinal field of absolute
value larger than about 200 -- 300 G, but is not biased with respect
to $v \sin i$. Any bias with respect to spectral type is primarily a
feature of the previous classification programmes that have identified
candidate magnetic Ap stars in clusters, rather than a feature imposed by
an instrumental field detection sensitivity that depends on the star's
spectroscopic features.

\subsection{The target list}\label{Sect_Target_List}
Two problems arise in the selection of a list of suitable targets for
a survey of magnetic fields in cluster stars: (a)
identification of candidate Ap stars in the field of individual
clusters (most of the known Ap stars are field stars) and
(b) determination of cluster membership.

Numerous studies have identified probable Ap members of clusters and
associations. Surveys for Ap/Bp cluster members based on
low-dispersion spectroscopy were reported by Hartoog (\cite{Har76},
\cite{Har77}), and by Abt (\cite{Abt79}). Probable Ap stars have also
been reported from a number of studies of individual clusters
(e.g. Dachs \& Kabus \cite{DacKab89}). Furthermore, the Michigan
Spectral Survey (e. g. Houk \& Smith-Moore \cite{HouSmi88}) has
provided uniform MK classifications for a large number of HD stars in
the fields of clusters south of $\delta = +5^\circ$. This allowed us
to select some Ap stars that are possible cluster members.

Another important source of identifications of cluster Ap stars has
been the use of photometric indices that are sensitive to a broad
depression near 5200 \AA\ in the energy distribution of most Ap stars
(e.g., Kupka, Paunzen, \& Maitzen \cite{Kupetal03}). In particular,
Maitzen and collaborators (e.g., Maitzen \cite{Mai93}) have developed
a narrow band photometric index ($\Delta a$) which is reasonably
sensitive to the spectral peculiarities of Ap stars with effective
temperatures in the range $\sim 8000 - 14\,000$\,K. Maitzen's group
has systematically obtained $\Delta a$ photometry of many clusters
to identify Ap stars. The $\Delta a$ system has also been used
by Joncas \& Borra (\cite{JonBor81}) to search for Ap stars in the
Orion OB1 association.

It is also known that the $Z$ index, which can be
computed for the many cluster stars for which Geneva photometry is
available, is a powerful discriminant of Ap stars in
approximately the same temperature range as the $\Delta a$ index 
(e. g. Kramer \& Maeder \cite{KraMae80}; Hauck \& North
\cite{HauNor82}).

We have made extensive use of the WEBDA cluster database
(e.g. Mermilliod \& Paunzen \cite{MerPau03}), Simbad, and a catalogue
of candidate cluster Ap stars by Renson (\cite{Ren92}).

Most of the surveys discussed above have made a serious effort to
determine cluster memberships, mainly on the basis of spatial location
and apparent magnitude. However, recent work based on and stimulated
by the Hipparcos astrometric space mission has led to a major
expansion in the available data on parallaxes and proper
motions. Hipparcos parallaxes (e.g. Gomez et al.\ \cite{Gometal98})
provide a valuable membership discriminant out to about 300 pc for
many cluster stars. Even more importantly, the Tycho (H{\o}g et al.\
\cite{Hogetal98}) and Tycho-2 (H{\o}g et al.\ \cite{Hogetal00}) proper
motion catalogues now provide powerful tests of membership
out to nearly one kpc, for a much larger number of stars, as
demonstrated for example by Robichon et al.\ (\cite{Robetal99}), de
Zeeuw et al.\ (\cite{deZetal99}), and Dias, L\'{e}pine, \& Alessi
(\cite{Diaetal01}).

Using the resources discussed above, a database containing a variety
of information on more than 200 suspected Ap cluster members in more
than 70 clusters and associations was constructed and used for the
selection of the targets. In general, we have given highest priority
to stars which appear to be probable Ap stars, and probable cluster
members, but we have also observed a number of stars for which at
least one of these criteria was uncertain. Up to the present time, we
have been able to observe (or occasionally find in the literature)
magnetic observations of about 1/3 of the stars in our database.

Cluster membership will be discussed in the second paper in this
series (Landstreet et al., in preparation, hereafter referred to as
Paper~II), where we will analyze the astrophysical and evolutionary
results of our survey in more detail. The present paper focuses on
the magnetic observations themselves.

\section{The basic formulae}\label{Sect_Basic}
The diagnostic tool for stellar magnetic fields used in this work
has been proposed by Bagnulo et al.~(\cite{Bagetal02}). Here we
review and elaborate this method.

The \textit{mean longitudinal magnetic field} \Bz, i.e., the
component of the magnetic field along the line of sight averaged over
the visible stellar disk, can be measured through the analysis of the
circular polarization of spectral lines. In the \textit{weak
field} regime (i.e., when the Zeeman splitting is small compared to
the line intrinsic broadening) we have (e.g. Landstreet \cite{Lan82})
\begin{equation}
\frac{V}{I} = - g_\mathrm{eff} \ \cz \ \lambda^{2} \
                \frac{1}{I} \
                \frac{\mathrm{d}I}{\mathrm{d}\lambda} \
                \Bz\;,
\label{EqBz}
\end{equation}
where $g_\mathrm{eff}$ is the effective Land\'{e} factor (1 for H
Balmer lines, see Casini \& Landi Degl'Innocenti \cite{CasLan94}),
\textit{V} is the Stokes parameter which measures the circular
polarization, \textit{I} is the total (unpolarized) intensity,
$\lambda$ is the wavelength expressed in \AA, \Bz\ is the longitudinal
field expressed in Gauss, and
\begin{equation}
\cz = \frac{e}{4 \pi m_\mathrm{e} c^2}
\ \ \ \ \ (\simeq 4.67 \times 10^{-13}\,\mathrm{\AA}^{-1}\ {\rm G}^{-1})
\end{equation}
where $e$ is the electron charge, $m_\mathrm{e}$ the electron mass, and
$c$ the speed of light. For a typical A-star atmosphere, the
weak-field approximation holds for field strength $\la 1$\,kG for
metal lines, and up to $\sim 10$\,kG for H lines.

A least-squares technique can be used to derive the longitudinal field
via Eq.~(\ref{EqBz}). We minimise the expression
\begin{equation}
\chi^2 = \sum_i \frac{(y_i - \Bz\,x_i - b)^2}{\sigma^2_i}
\label{EqChiSquare}
\end{equation}
where, for each spectral point $i$, $y_i = (V/I)_i$, $x_i =
-g_\mathrm{eff} \cz \lambda^2_i (1/I\ \times \mathrm{d}I/\mathrm{d}\lambda)_i$, and $b$
is a constant. For each spectral point $i$, the derivative
of Stokes~\textit{I} with respect to wavelength is evaluated as
\begin{equation}
\left(\frac{\mathrm{d}I}{\mathrm{d}\lambda}\right)_{\lambda = \lambda_{i}} =
\frac{\mathcal{N}_{i+1} - \mathcal{N}_{i-1}}{\lambda_{i+1} - \lambda_{i-1}}
\label{EqDerivative}
\end{equation}
where $\mathcal{N}_i$ is the photon count at wavelength
$\lambda_i$. The estimates for the errors on the parameters \Bz\ and
$b$ are given by the diagonal elements of the inverse of the $\chi^2$
matrix (see, e.g., Bevington \cite{Bev69}).

The application of a least-squares technique is justified if
\begin{equation}
\Bz\ \ \sigma_{x_i} \ll \sigma_{y_i}
\label{EqSigmax}
\end{equation}
(see, e.g., Bevington \cite{Bev69}), where 
\begin{equation}
\sigma^2_{x_i} = 
\left(-g_\mathrm{eff}\, \cz\, \lambda^2_i\right)^2 \
\left( \left(-\frac{1}{I^2}\right)^2\,
       \left(\frac{\mathrm{d}I}{\mathrm{d}\lambda_i}\right)^2\,\sigma^2_I 
      + \left( \frac{1}{I} \right)^2\, \sigma^2_{{\rm d}I/{\rm d}\lambda} \right)
\end{equation}
and
\[
\sigma^2_{y_i} = \sigma^2\left[(V/I)_i\right]
\]
For the stars studied in this work, we have verified that, at all
observed wavelengths, Eq.~(\ref{EqSigmax}) is verified.

\section{Instrument and instrument setup}\label{Sect_Instrument}
The Focal Reducer/Low Dispersion Spectrograph FORS1 of the ESO VLT is
a multi-mode instrument equipped with polarimetric optics.
The instrument characteristics and performances are outlined by Seifert et al.\
(\cite{Seietal00}) and in the FORS1/2 User Manual
(VLT-MAN-ESO-13100-1543). In this survey FORS1 has been used as a low
resolution spectropolarimeter.

\subsection{Instrument mode}\label{Sect_Ins_Mode}
Spectropolarimetry with FORS1 can be performed in two different
observing modes: \textit{fast mode} and \textit{fims
mode}. \textit{Fast mode} is used for observations of individual
objects that can be acquired `fast' with a simple centering on the
slit. \textit{Fims} mode permits one to place slitlets on up to nine
targets within a $6.8' \times 6.8'$ field of view
(multi-object spectropolarimetry). Whenever our main
targets were in a field including other stars of similar magnitude,
we observed in \textit{fims mode}. In this way, in addition to
polarized spectra of Ap stars, we obtained polarized spectra of
numerous normal A and B-type cluster stars.

\subsection{Grism choice}
We have used the grisms 600\,B and 600\,R, which cover the spectral
ranges 3450--5900\,\AA, and 5250--7450\,\AA, respectively. Grism
600\,B covers all H Balmer lines from H$\beta$ down to the Balmer
jump; grism 600\,R covers H$\alpha$. Both grisms have 600
grooves\,mm$^{-1}$, and yield dispersions of 1.20\,\AA, and 1.08\,\AA\
per pixel, respectively.  With a 1\arcsec\ slit width they provide
spectral resolving powers of 780 and 1160, respectively.  

For this study, grism 600\,B seems a better choice than 600\,R.
Although H$\alpha$ is more sensitive to the magnetic field
than individual H Balmer lines at shorter wavelengths (the Zeeman
effect depends quadratically upon wavelength), the combined analysis
of several Balmer lines from H$\beta$ down to the Balmer jump leads to
a smaller error bar than the analysis of H$\alpha$ only.  Furthermore,
the flux of an A-type stars is larger in the blue than in the red,
therefore the same SNR can be reached with a shorter integration time
using grism 600\,B rather than grism 600\,R. On the other hand, it
should be noted that the useful field of view in \textit{fims mode}
depends on the spectral range that one wishes to cover: the larger the
spectral range that one needs to observe, the smaller the spatial
region where one can place slitlets. Hence, in \textit{fims mode},
grism 600\,R offers more flexibility than grism 600\,B, as its useful
wavelength coverage corresponds to H$\alpha$ only, which does not put
a strong constraint on the field of view.  In (ESO period)
P68 we used grism 600\,R (in combination with the order
separation filter GG\,435). In P70, P72, P73, and P74, we 
used grism 600\,B (with no order separation filter).

For most of the observations, we have used a slit width of 0.5\arcsec
or 0.8\arcsec.

\subsection{CCD readout}
In polarimetric mode, overheads represent a significant fraction of
the total telescope time. Ultra-high SNR observations (needed to
reach the wanted uncertainty of $\sim 100$\,G) require multiple exposures,
hence, multiple CCD readouts. We minimized the CCD readout time as
follows. In case of observations of individual stars (\textit{fast
mode}) we used a windowed read-out mode of $2048 \times 400$
pixels. In case of multi-object observations (\textit{fims mode})
where the full size of the CCD was needed, we used a 4-port readout.
The lowest available gain was selected, as to maximise the actual
electron capacity, hence, the SNR that can be reached with each single
exposure.

\section{Data reduction}\label{Sect_Data}
\subsection{Frame pre-processing, spectrum extraction and wavelength 
            calibration}\label{Sect_Prepro} 

The data have been reduced and optimally extracted using standard IRAF
routines.  All the science frames have been bias subtracted with the
corresponding master bias obtained from a series of five frames taken
the morning after the observations.  No flat fielding procedure has
been applied to our data. By performing reduction experiments with and
without flat-fielding, we have verified that flat fielding does not
influence significantly the final computation of the Stokes
profiles. In fact, flat fields obtained with the grism 600\,B are
severely affected by internal reflections from the Longitudinal
Atmospheric Dispersion Corrector (LADC). Frames obtained in
\textit{fims mode} were read out in four ports.  The CCD is thus
divided in four quadrants, each of them characterized by its own bias
level and gain. In order to compensate for the different gains, we
multiplied each science frames by the ratio of an imaging screen flat
read out in one port and an imaging screen flat read out out in four
ports.

When extracting the spectra we found that the use of standard
extraction apertures ($\sim 10$ pixels width) introduced artefacts
into the Stokes~$V$ spectrum.  This problem was solved using
apertures as large as $\sim 4-5$ times the spatial FWHM of the
spectrum, i.e., typically 50 pixels width.  Apertures that are not symmetric
about the flux peak have been used for stars close to the edge of the
slit. This occurred occasionally for targets observed in \textit{fims
mode}.  We used a high-order ($\sim 15$) Legendre fitting function to
trace the spectrum\footnote{The high order of the fitting function is
justified by the high SNR that makes it possible to trace very
accurately the spectrum across the CCD.}.  Lower order functions were
used only in the case of low SNR secondary targets (typically in
frames obtained in \textit{fims mode}) and/or in the case of spectra
the length of which occupied just a fraction of the CCD (again in
\textit{fims} frames, depending on the positioning of the slitlets).

Sky subtraction was performed differently for spectra obtained in
\textit{fast} and \textit{fims mode}. In the first case the sky
subtraction was performed selecting symmetric regions on the left and
right side of each spectrum (typically between pixel 40 and 50 from
the central peak), and fitting those with a Chebyshev polynomial.  In
the case of data obtained in \textit{fims mode}, whenever the star was
not positioned at the center of the slitlet, the sky was estimated on
just one region at one side of the spectrum. In fact, we found that
sky-subtraction is not critical in the sense that it does not
significantly affect the final results. In some cases, we preferred
not to perform sky-subtraction at all, because of the presence of
LADC reflections close to the spectrum.

The FORS1 calibration plan includes wavelength calibrations frames
obtained at all retarder waveplate positions used for the science.
However, we found that the best and safest strategy is to use, for a
complete set of science data, just a single wavelength calibration
frame, and not match science and wavelength calibration frames
according to their retarder waveplate angles. We found that this
latter method occasionally introduces spurious polarization signals.

Wavelength calibration typically led to RMS scatter of $\sim
0.1$\,pixels and maximum error of $\leq 20$\,km\,s$^{-1}$. The fine
tuning of wavelength calibration based on night sky lines could not be
performed. Therefore the accuracy of the wavelength calibration is
restricted by instrument flexures, which are expected to be less than
1 pixel up to a zenith distance of 60\degr (see FORS1/2 User Manual).
Numerical tests show that this is of negligible impact on the
determination of the mean longitudinal magnetic field, using the
method described by Bagnulo et al.\ (\cite{Bagetal02}) and elaborated
below.

\subsection{Obtaining Stokes~$V$}
Circular polarization measurements are performed by inserting the
quarter-wave plate and the Wollaston prism into the optical path. A
combination of exposures taken at different waveplate orientations
allows one to minimise the contributions of spurious (instrumental)
polarization. The FORS user manual explains that instrumental polarization
cancels out to the first order if Stokes~$V$ is obtained from
\begin{equation}
\frac{V}{I} = \frac{1}{2}  \Bigg\{  
         \left(\frac{\fo - \fe}{\fo + \fe}\right)_{\alpha = -45\degr} -  
         \left(\frac{\fo - \fe}{\fo + \fe}\right)_{\alpha = +45\degr}
                           \Bigg\} \;,
\label{StokesUno}
\end{equation}
where \fo\ and \fe\ are the ordinary and extraordinary beams,
respectively, and $\alpha$ is the position angle of the
retarder waveplate (in fact, note that the FORS1/2 manual gives the
formula with the opposite sign). 
Another possibility is to obtain
Stokes~$V$ from
\begin{equation}
\frac{V}{I} = \frac{r-1}{r+1}
\label{StokesDue}
\end{equation}
where
\[
r^2 = \frac{\left(\fo/\fe\right)_{\alpha =-45\degr}}
           {\left(\fo/\fe\right)_{\alpha =+45\degr}}
\]
(e.g., Donati et al.\ \cite{Donetal97}).
We have verified that Eqs.~(\ref{StokesUno}) and (\ref{StokesDue}) give
consistent results both for Stokes~$V$ and \Bz.
 
The error bar associated with the Stokes~$V/I$, computed via
Eq.~(\ref{StokesUno}) is
\begin{equation}
\begin{array}{rcl}
\sigma^2\left[(V/I)\right] & = &
  \left(\left(\frac{\fe}{(\Den)^2}\right)^2 \sigma^2_{\fo} +
        \left(\frac{\fo}{(\Den)^2}\right)^2 \sigma^2_{\fe}\right)_{\alpha = -45\degr} + \\
             &   &
  \left(\left(\frac{\fe}{(\Den)^2}\right)^2 \sigma^2_{\fo} +
        \left(\frac{\fo}{(\Den)^2}\right)^2 \sigma^2_{\fe}\right)_{\alpha = +45\degr} \\
\end{array}
\label{sigmaV}
\end{equation}
Equation~(\ref{sigmaV}) can be simplified as follows. Let us assume
\[
\left(\sigma^2_{\fo}\right)_{\alpha = \pm 45\degr} = 
\left(\sigma^2_{\fe}\right)_{\alpha = \pm 45\degr} = 
\left(\fo\right)_{\alpha = \pm 45\degr} = 
\left(\fe\right)_{\alpha = \pm 45\degr} = \mathcal{N}
\]
where $\mathcal{N}$ is an estimate of the photon count in a given wavelength
range. 
For $m$ pairs of observations (i.e., one exposure at $\alpha = 45\degr$,
and one exposure at $\alpha = -45\degr$, repeated $m$ times), 
Eq.~(\ref{sigmaV}) becomes
\begin{equation}
\sigma^2\left[(V/I)\right] = \frac{1}{4m\,\mathcal{N}}
\label{EqSNR}
\end{equation}
Equation~(\ref{EqSNR}) simply states that the error
bar on $V/I$ decreases as $1/\sqrt{\mathcal{N}}$. The
polarization in a given wavelength range can be measured with an
0.1\,\% uncertainty if at least $2.5\times 10^5$ photons are accumulated
both in the ordinary and in the extraordinary beam, and in each of
the two exposures of a pair.

In order to detect weak magnetic fields ($\la 300$\,G) with the
technique used in this work, one has to obtain ultra high SNR ($\ga
1500$\,\AA$^{-1}$) observations. Even with a 8\,m telescope, this can be
achieved only on relatively bright stars ($V \la 13$, if we limit the
shutter time to $\la 2$\,h). Due to the limited CCD well capacity, multiple
exposures have to be taken. From a practical point of view, one has
to set the exposure time to a value that maximises the photon count
without risk of CCD saturation (e.g., by adjusting the exposure time to
get a peak count of 30\,000 ADU per pixel), and then take several
pairs of exposures with the retarder waveplate at the $+45\degr$ and
$-45\degr$ positions. Equation~(\ref{EqSNR}) can be explicitly
expressed in terms of ADU in the following way. Let us define
$M$ as the ADU per pixel and $g$ as the number of electrons per
ADU, so that actual photon count $\mathcal{N}$ is given by $gM$. Let
us also define $\mathcal{A}_\mathrm{s}$ as the ratio between the ADU
integrated in a pixel column along the direction perpendicular to the
dispersion, and the peak ADU in the central
pixel. The error bar on the circular polarization measured in the
wavelength interval $\Delta \lambda$ covered by 1 pixel is given by
\[
\sigma^2\left[(V/I)\right](\Delta \lambda) = 
\frac{1}{4\,m\,g\,\mathcal{A}_\mathrm{s}\,M_{\rm max}}
\]
where $M_{\rm max}$ is the peak ADU.

Recalling the properties of a Gaussian, we can write
$\mathcal{A}_\mathrm{s} =1.065$\,FWHM.  With a plate scale of
0.2\arcsec\ per pixel, as in the case of FORS1, and with 0.8\arcsec\
seeing, $\mathcal{A}_\mathrm{s} \simeq 4$.  Assuming $g=2.9$ (a
typical value for the FORS1 ``low gain'' readout mode), and setting as
a peak value $M_{\rm max}=30\,000$\,ADU, we get
\begin{equation}
\sigma\left[(V/I)\right](\Delta \lambda) \simeq 8 \times 10^{-4}\ \frac{1}{\sqrt{m}}\;.
\label{Eq_Practical}
\end{equation}
Equation~(\ref{Eq_Practical}) allows one to calculate the number of
pairs of exposures that are needed to measure $V/I$ with a given error
bar.
For instance,
with a pair of exposures at half the CCD full well, one gets an
error bar close to 0.1\,\%. To go below the threshold of an 0.01\,\%
error bar (if possible at all) one should obtain 64 pairs of 
exposures.

We found that measuring $V/I$ with an accuracy of a few units in
$10^{-4}$ per \AA\ in the continuum near to H$\beta$ allowed us to
measure magnetic fields with an error bar between 50 and
100\,G. Therefore, we decided that our observing strategy would be
based on a series of four pairs of exposures (following the sequence
$\alpha = +45\degr$, $-45\degr$, $-45\degr$, $+45\degr$,
etc.). However, we could limit the number of pairs of exposure to four
only when we obtained telescope time in visitor mode, which allowed us
to optimize the exposure time based on the weather conditions. During
P72 and P73 we were allocated telescope time in service
mode, and were forced to set the exposure time to
conservatively low values to be sure to avoid CCD saturation. In
these cases the number of pairs of exposures was increased from four
to six or eight to guarantee a sufficient SNR.

Thought must be given to the ratio between shutter time and overhead
time, as the latter ($\sim 20$ minutes for a series of 4 pairs of
exposures) may represent a substantial fraction of the total 
time for a single pointing. 
Using grism 600\,B with a 0.5\arcsec\ slit width, we can obtain
a peak ADU count of $\sim 30\,000$ in about 10 minutes shutter time
for a $V=11-12$ A-type star (depending on the weather conditions), and
in just 1 minute for $V=8.5-9.5$.

\subsection{The use of a $\sigma$-clipping algorithm}
\label{SecSigmaClipping}
The simplest method to obtain $V/I$ is to add up all spectra
obtained in the same beam and with same retarder waveplate position,
and then use Eq.~(\ref{StokesUno}). However, reduction products can
be improved by using a $\sigma$-clipping algorithm as follows.
For each wavelength step one calculates
\begin{equation}
\begin{array}{l}
\left(\frac{V}{I}\right)_{ij} =
\frac{1}{2}  \Bigg\{
\left(\frac{\fo_i - \fe_i}{\fo_i + \fe_i}\right)_{\alpha = -45^\circ } -
\left(\frac{\fo_j - \fe_j}{\fo_j + \fe_j}\right)_{\alpha = +45^\circ }
\Bigg\}\\
\ \ i,j=1,2,\ldots,m\\
\end{array}
\label{EqVij}
\end{equation}
where $m$ is the total number of pairs of observations. Then, one
calculates the median $\widehat{(V/I)}$ from the $m^2$ $(V/I)_{ij}$
values, and the median absolute deviation (MAD), i.e., the median of
the distribution
\[
\vert \left( (V/I)_{ij} - \widehat{(V/I)} \right) \vert \;.
\]
Setting $\sigma = 1.48$\,MAD (e.g., Huber \cite{Huber81},
pp.~107--108), those $(V/I)_{ij}$ values for which
\[
\vert (V/I)_{ij} - \widehat{(V/I)} \vert > \mathcal{K} \sigma
\]
are rejected (we typically adopted $\mathcal{K} = 3$). The procedure
is iterated until no points are rejected, but from the second
iteration on, the median $\widehat{(V/I)}$ is replaced by the weighted
average
\begin{equation}
    \overline{(V/I)} = 
    \sum_{ij} (V/I)_{ij}\; \sigma^{-2}[(V/I)_{ij}]\; /\; 
    \sum_{ij} \sigma^{-2}[(V/I)_{ij}] \; ,
    \label{EqFinalV}
\end{equation}
where the sum is obviously extended over the $K$ $(V/I)_{ij}$ values
that have not been rejected by the $\sigma$-clipping algorithm ($K \le
m^2$), and $\sigma^2\left[(V/I)_{ij}\right]$ is given by Eq.~(\ref{sigmaV}).

If no value in the $m^2$ set of $(V/I)_{ij}$ pairs has been rejected, and
if we assume that the errors given by Eq.~(\ref{sigmaV}) are approximately
equal for all spectra, we can estimate the error bar of $\overline{(V/I)}$
as:
\begin{equation}
  \label{EqPhotonNoise}
  \left\{\sigma^2\left[\overline{(V/I)}\right]\right\}' \simeq 
  \frac{1}{m}\sigma^2\left[(V/I)\right] \; .
\end{equation}

An alternative estimate of the same error bar can be obtained as follows:
\begin{equation}
  \label{EqStatistics}
  \left\{\sigma^2\left[\overline{(V/I)}\right]\right\}'' \simeq 
  \frac{1}{m^2(m-1)} \sum_{ij} [(V/I)_{ij} - \overline{(V/I)}]^2 \; .
\end{equation}

In practice, at each
wavelength step, we adopt the error bar given by the maximum among the
values obtained from Eq.~(\ref{EqPhotonNoise}) and
Eq.~(\ref{EqStatistics}).
The above two expressions will be derived in detail in
Appendix~\ref{app:sigma_v}, together with more general expressions
valid for the case of one or more exposures rejected by the
$\sigma$-clipping algorithm.


\subsection{Measuring \Bz}\label{Sect_Bz_Meas}
The mean longitudinal field \Bz\ is obtained as explained in
Sect.~\ref{Sect_Basic}, using $V/I$ obtained via
Eq.~(\ref{EqFinalV}). Following common statistics guidelines, one
should consider a detection as `definite' whenever the relation
\begin{equation}
\vert \Bz  \vert / \sigma\left[\Bz\right]  \ge 3 \\
\label{Eq_Detection}
\end{equation}
is satisfied.

We encountered a number of cases for which the field was detected at
about 3-$\sigma$ level, and where minor changes in the data reduction
would transform a marginal detection in a null or in a definite
detection. Although these cases should certainly be investigated via
additional observations, we tried to extract further information from
the available spectra, to formulate a more robust and
reliable criterion for field detection.

First, we decided to explore an alternative method for the
determination of the mean longitudinal field.  From the individual
pairs of $V_{ij}/I_{ij}$ and $I_{ij}$ given by
Eqs.~(\ref{EqVij}), we calculated $m^2$ $\Bz_{ij}$\ values, and
the weighted mean longitudinal field
\begin{equation}
\Bz'\ = 
\sum_{ij} \Bz_{ij}\, \sigma^{-2}\left[\Bz_{ij}\right] /
\sum_{ij} \sigma^{-2}\left[\Bz_{ij}\right]
\;.
\label{EqBzprime}
\end{equation}
The corresponding error bar $\sigma\left[\Bz'\right]$ was calculated as
\[
\sigma^2\left[\Bz'\right] = \frac{\sum_{ij=1}^m (\Bz_{ij} - \Bz')^2}{m^2 (m - 1)}
\]
For each star, we systematically checked the consistency
between the \Bz\ value determined through the average $V/I$
obtained via Eq.~(\ref{EqFinalV}) and the $\Bz'$\ value
obtained via Eq.~(\ref{EqBzprime}). We also checked whether
the relation
\begin{equation}
\vert \Bz'\ \vert / \sigma\left[\Bz'\right] \ge 3 
\label{Eq_Detection_Prime}
\end{equation}
was satisfied.

Second, we performed a systematic analysis of metal lines.  As pointed
out earlier, Eq.~(\ref{EqBz}) is formally valid only under the
weak-field approximation. Therefore, in principle, \Bz\ measurements
by our method should be performed only on H Balmer lines. Furthermore,
H Balmer lines are well sampled even at the low resolution of our
observations, whereas metal lines are unresolved. Nevertheless, we
found that the \Bz\ determined via metal line analysis is consistent
with that measured from H Balmer lines, provided that \Bz\ is $\la
800$\,G (see Sect.~\ref{Sect_weak}). Therefore we decided to analyze
the metal lines, i.e., to apply the least-squares
technique to spectral regions free from H Balmer lines. In
addition, we determined \Bz\ using the whole spectrum, i.e., including
both Balmer and metal lines. The outcome of this analysis will be
discussed in Sect.~\ref{Sect_Balmer_Metal}.

\section{Magnetic field determinations from Balmer lines vs. 
field determinations from metal lines}\label{Sect_Balmer_Metal}
\subsection{The magnetic ``standard'' star HD~94660}\label{Sect_HD94660} 
\begin{table*}
\input{4223table1.tex}
\end{table*}
In order to compare our results with those obtained through different
techniques, we repeatedly observed a well known magnetic Ap
star: HD~94660 (= KQ~Vel).  Previous \Bz\ measurements of HD~94660 were
obtained using the H$\alpha$ Balmer line (Borra \& Landstreet
\cite{BorLan75}), the H$\beta$ Balmer line (Bohlender et al.\
\cite{Bohetal93}), and metallic lines (Mathys \cite{Mathys94}; Mathys
\& Hubrig \cite{MatHub97}). Moreover, HD~94660 is the star observed
by Bagnulo et al.\ (\cite{Bagetal02}) to develop the technique used in
this work.

\begin{figure}
\includegraphics*[width=9cm]{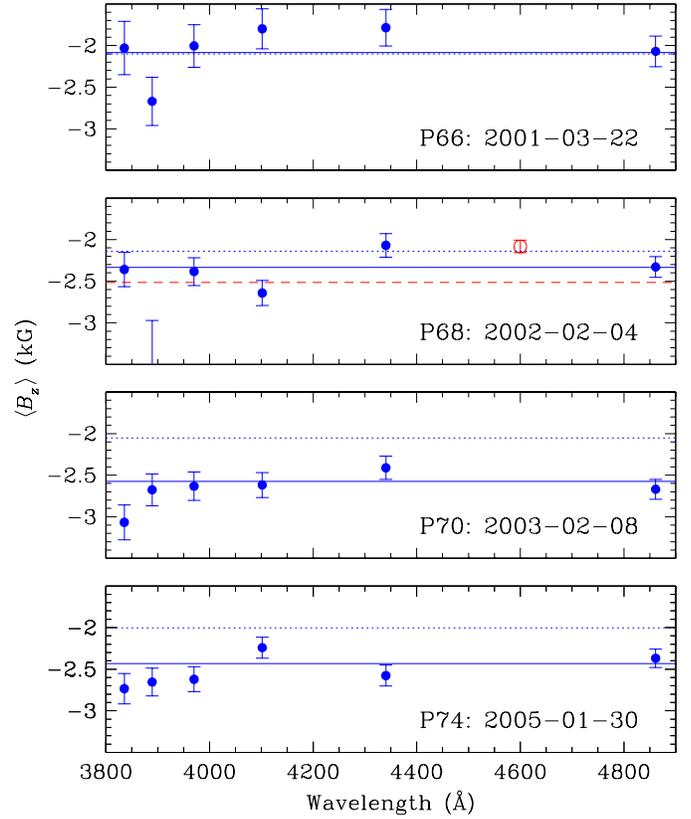}
\caption{\label{FigBalmers} \Bz\ measurements of HD~94660 from
individual Balmer lines. Each panel refers to a different observing
date as specified in the figure. The \Bz\ values measured from
H$\beta$ and blueward Balmer lines are plotted with filled circles at
the corresponding wavelength. The \Bz\ determination from H$\alpha$
obtained on 2002-02-04 is arbitrarily plotted with an empty circle at
$\lambda=4600$\,\AA. The solid lines correspond to \Bz\
obtained using all Balmer lines from H$\beta$ down to the Balmer jump,
and the dotted lines indicate \Bz\ obtained from the metal lines
with grism 600\,B. The dashed line in the second panel shows the \Bz\
value obtained from the metal lines observed with grism 600\,R.}
\end{figure}

\begin{figure}
\includegraphics*[width=9cm]{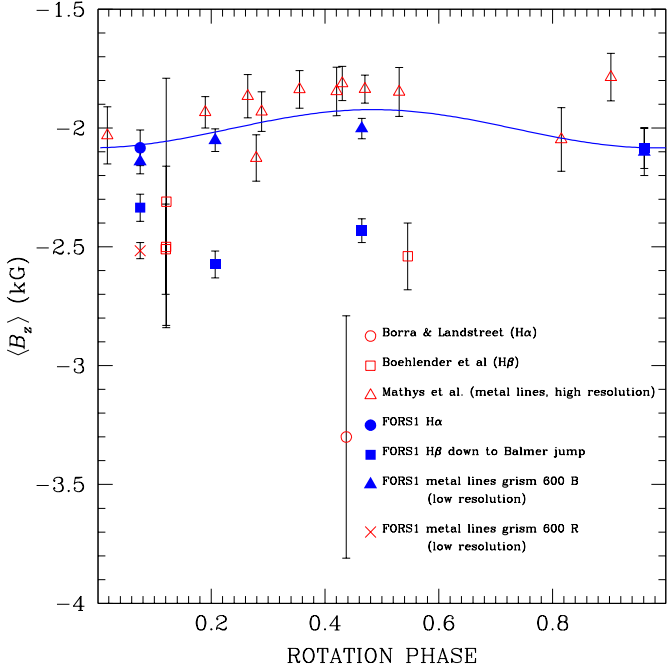}
\caption{\label{FigHD94660} Longitudinal field determinations of
HD~94660.  The solid line is a fit to the \Bz\ determinations from
metal lines.  We have used a first-order Fourier expansion. At phase
0.075 we have considered the \Bz\ determination obtained with grism
600\,B and not the one obtained with grism 600\,R. For the star's
rotation period we adopt 2800\,d (Landstreet \& Mathys
\cite{LanMat00}), and the zero phase point is at JD = 2446500.0.}
\end{figure}

The log of our observations and \Bz\ measurements is given in
Table~\ref{Table_HD94660}. Note that on 2002-02-04 we obtained two
consecutive \Bz\ measurements: the first one with grism 600\,B, the
second one with grism 600\,R. 

Figure~\ref{FigBalmers} shows the \Bz\ measurements obtained from both
the Balmer and the metal lines. The \Bz\ values obtained from the
Balmer lines blueward of H$\beta$ are consistent among themselves, and
marginally consistent with the \Bz\ value obtained from
H$\alpha$. With the exception of the measurement taken on 2001-03-22,
the \Bz\ values obtained from the metal lines are not consistent with
those obtained from Balmer lines. This is probably due to the fact
that Eq.~(\ref{EqBz}) is valid only under the weak field approximation
(which in this case is correct for Balmer lines but not for metal
lines). Note also that the \Bz\ values measured via metal line
analysis are not consistent among themselves if we compare data
obtained with grism 600\,B and grism 600\,R. The comparison between
\Bz\ values obtained from H Balmer lines and metal lines will be
further discussed in Sect.~\ref{Sect_weak}.

We also compared our \Bz\ determinations of Table~\ref{Table_HD94660}
with those previously obtained in the literature, adopting for the
star's rotation period 2800\,d (Landstreet \& Mathys \cite{LanMat00}).
The results are shown in Fig.~\ref{FigHD94660}. In general, it appears
that \Bz\ values obtained from Balmer lines are \textit{not}
consistent with the \Bz\ determinations obtained using metallic lines.
It seems likely that different methods used to evaluate \Bz\ bear
systematic differences (even though each may be internally
consistent).  Systematic inconsistencies between \Bz\ determinations
in Ap stars obtained with different chemical elements or with
different techniques have been already found in previous works (see,
for instance, Ryabchikova et al.\
\cite{Ryaetal05}, who analysed several observations of HD~24712).

\subsection{Weak field Ap stars}\label{Sect_weak}
Using all our data collected for Ap stars, we show here that the
metal line analysis and the Balmer line analysis produce consistent
results when the weak field hypothesis is satisfied.

\begin{figure}
\includegraphics*[width=9.0cm]{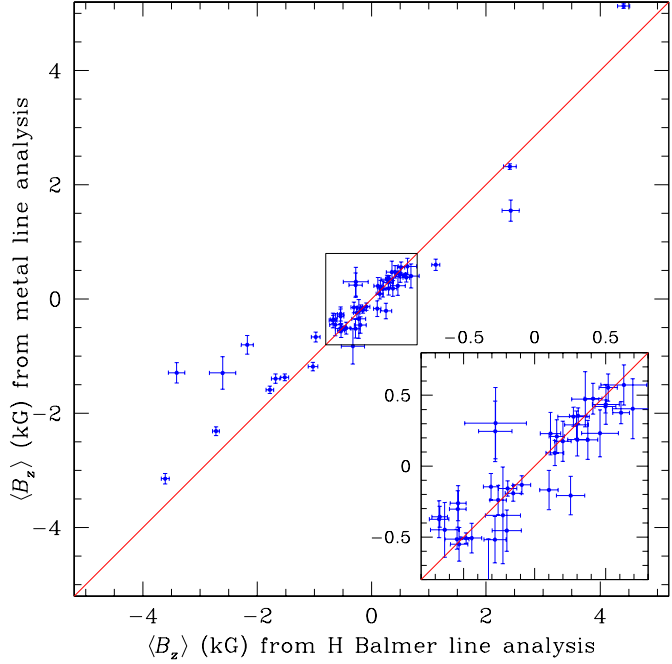}
\caption{\label{FigBalmetal} \Bz\ determinations from metal lines
versus \Bz\ determinations from Balmer lines for the stars of
Table~\ref{Table_Peculiars}. The solid line has slope=1. For clarity,
only stars for which \Bz\ obtained from Balmer lines is such that
$\vert\Bz\vert / \sigma[\Bz] \ge 1$ have been plotted. The outliers
HD~318107 and NGC~2244\,334 (see Table~\ref{Table_Peculiars}) are not
plotted.}
\end{figure}

Figure~\ref{FigBalmetal} shows the \Bz\ values obtained from metal
lines versus those obtained from Balmer lines for the observed Ap
stars. It appears that the two methods give consistent \Bz\ values for
$\vert\Bz\vert \la 800$\,G.  Above the 1\,kG level, differences
between Balmer line and metal line technique become noticeable or even
striking.  E.g., for HD 310187 we obtained from the Balmer lines $\Bz
= 6519 \pm 55$\,G, and from the metal lines we obtained $\Bz = 3784
\pm 59$\,G. In general, above the 1 kG level, the modulus of \Bz\
determined from Balmer lines is larger than that from the metal lines,
as we would expect from the earlier breakdown of the weak field
expression for metal lines.

\begin{figure}
\includegraphics*[angle=90,width=9.0cm]{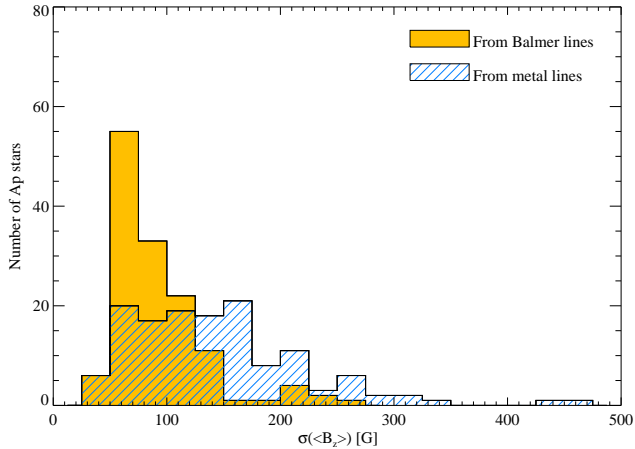}
\caption{\label{FigHistoerror} Histograms of the error bars for the
\Bz\ measurements of Table~\ref{Table_Peculiars} obtained from Balmer
lines and from metal lines. The median of the distribution of the
error from Balmer lines is 78\,G, from metal lines it is 130\,G.}
\end{figure}

Summarizing, it appears that if the field is weak, both methods are
consistent, and each gives an indication whether the star is magnetic
or not \textit{independent} of the other.

Figure~\ref{FigHistoerror} compares the distributions of the error
bars for \Bz\ obtained via Balmer and metal line analysis for the
observed Ap stars. We see that the distribution of the error bars
calculated via metal line analysis is broader than that obtained from
Balmer lines. This is due primarily to the fact that Balmer lines have
similar strength in all A and B-type stars, whereas metal lines may
change greatly from star to star.  For a given SNR, the error
bars obtained via metal line analysis are smaller in spectra that are
richer in metal lines than in spectra that are poorer in metal
lines. As expected, for a set of observations of similar SNR, the
Balmer line analysis leads to results characterised by more
homogeneous error bars than the metal line analysis.

\section{Observations}\label{Sect_Observations}
\begin{table}
\input{4223table2.tex}

\end{table}
\begin{table*}
\input{4223table3.tex}
\end{table*}
\begin{figure}
\includegraphics*[width=9cm]{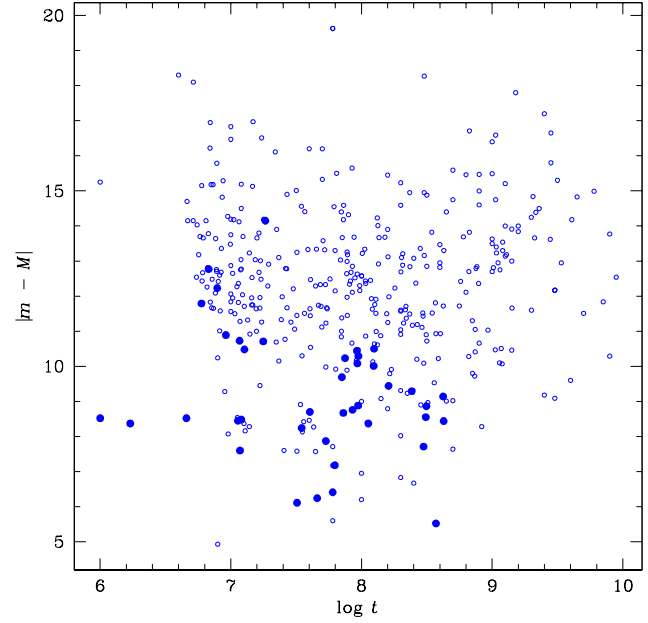}
\caption{\label{Fig_Clusters} Distance modulus versus age of open
clusters. The empty circles show all open clusters with known age
and $\delta <+10\degr$. Filled circles are the clusters observed in
this survey (same data as in Table~\ref{Table_Cluster_List}). Data
have been extracted from WEBDA.}
\end{figure}

In this survey we have observed 97 Ap stars, 138 normal A and B-type
stars, and 22 non early-type stars. All these stars are candidate
members of the open clusters (or the associations) listed in
Table~\ref{Table_Cluster_List}.  Figure~\ref{Fig_Clusters} gives an
overview of the range in age and distance modulus of the target
clusters, compared to all Milky Way open clusters with known age 
and $\delta \le +10^\circ$. A summary of the observing log is given
in Table~\ref{Table_Log}.

Target Ap stars have been selected as explained in 
Sect.~\ref{Sect_Target_List}, and are listed in
Table~\ref{Table_Peculiar_Names}, together with the $V$ magnitude and
the spectral type extracted from the \textit{General Catalogue of Ap
and Am stars} (Renson, Gerbaldi, \& Catalano \cite{Renetal91}), and
from SIMBAD. Searching the literature (in particular the catalogue by
Bychkov, Bychkova, \& Madej \cite{Bycetal03}), we found that 10 of
these Ap stars have already been checked for magnetic field in
previous studies. We observed some of our objects more than once 
(we have made 121 observations of Ap stars). In total we
obtained 52 field detections in 41 different stars (although for five
stars we have only marginal detections). Thirty-six Ap stars in which
we have detected a field were not previously known as magnetic. A few
Ap stars of Table~\ref{Table_Peculiar_Names} are in fact
\textit{cluster non-members}. Chemical peculiarity, and, above all,
cluster membership, will be further discussed in Paper~II, where we
will also provide estimates for star's temperature, mass, and
evolutionary state. Two remarkable stars have been the subject of
previous papers: HD~66318 (Bagnulo et al.\ \cite{Bagetal03}) and NGC
2244~338 (Bagnulo et al.\ \cite{Bagetal04}).

The normal A and B-type stars observed in this survey are listed in
Table~\ref{Table_Normal_Names}, together with their $V$ magnitudes and
spectral classifications. In many cases the spectral type was not
available in SIMBAD or in the literature, therefore, it has been
determined through their color indices $B-V$ and $U-B$ extracted from
WEBDA.  The observed color indices have been corrected for extinction
using the cluster $E(B-V)$ (also reported in WEBDA) and
applying the extinction law by Cardelli, Clayton, \& Mathis
(\cite{Caretal89}). The dereddened colours have been then compared to
those of typical main sequence stars to determine the approximate
spectral type of each object. FORS1 spectra were used to roughly check
the validity of this spectral classification, but no attempt was made
to refine it. The spectral type obtained from the colour indices are
reported between squared brackets, and should be used with caution.  We
discovered {\it a posteriori} that the spectra of $\sim 20$ normal
stars were in fact not of early-type. These cases are listed in
Table~\ref{Table_NonA}.  We do not report observations of a few
additional targets which have spectra with a too low SNR to be useful
for magnetic studies.

In this work we report the \Bz\ values obtained from the analysis of
both Balmer lines and metal lines.  For peculiar stars, the medians of
the errors obtained from the Balmer lines and from the metal lines are
78\,G, and 130\,G, respectively. For normal stars they are 136\,G, and
173\,G, respectively.  For normal stars the median error bar is larger
than for peculiar stars because we optimized integration times to
reach the highest possible SNR for Ap stars, rather than for the
secondary targets added in the \textit{fims mode} observations.  In
addition, we calculated the field measured from the analysis of the
whole spectrum, including metal lines. 
The latter method formally gives \Bz\ values with the
smallest error bar: for Ap stars the median is 62\,G, and for
normal stars is 94\,G. Although some caution is required in
interpreting a \Bz\ determination obtained combining Balmer and metal
lines, one can reasonably argue that this method may be used to
reinforce the conclusions achieved with the two primary methods.

The \Bz\ determinations for Ap stars and for normal A and B stars are
reported in Tables~\ref{Table_Peculiars} and \ref{Table_Normals},
respectively. These Tables are organized as follows. The day and time
of the observations are given in col.~2, and 3, respectively, and
correspond to the midpoint between the instant when the shutter opens
for the first exposure and the time when the shutter closes after the
last exposure. Columns~4, 5, and 6 report the \Bz\ values obtained
from the analysis of the Balmer lines, the metal lines, and the whole
spectrum, respectively. All these \Bz\ values are obtained through the
least-squares technique applied to the averaged $V/I$ calculated via
Eq.~(\ref{EqFinalV}). For each method, we also systematically
calculated $\Bz'$ from Eq.~(\ref{EqBzprime}). The cases in which both
Eqs.~(\ref{Eq_Detection}) and (\ref{Eq_Detection_Prime}) are satisfied
are indicated in col.~7 with symbol ``D''. The cases where only one of
these Eqs.\ is satisfied are flagged with symbol ``d''. If neither of
Eqs.~(\ref{Eq_Detection}) or (\ref{Eq_Detection_Prime}) is satisfied,
we use the symbol ``n''. For example, in HD~62992, using the average
spectrum obtained via Eq.~(\ref{EqFinalV}), we have measured from
Balmer lines $\Bz = -190 \pm 65$\,G.  This value (slightly less than a
3-$\sigma$ detection) is reported in col.~4 of
Table~\ref{Table_Peculiars}.  From the average of the \Bz\ values
obtained from the individual spectra we obtain $\Bz' = -186 \pm 54$\,G
(not reported in Table~\ref{Table_Peculiars}). This is above a
3-$\sigma$ detection, hence to this \Bz\ determination we associate
the symbol ``d''.

This procedure is repeated for each method used (Balmer lines, metal
lines, full spectrum), so that to each \Bz\ value, a three-term flag
is associated. This three-terms flag is meant to help to evaluate the
significance of each field detection. A ``DDD'' sequence clearly
indicates a firm field detection, and ``nnn'' sequence indicates a
null detection, whereas intermediate cases deserve further
investigation and discussion. In general, a ``DnD'' or ``Dnd''
sequence may still represent field detection, if the lack of a
detection from the analysis of the metal lines can be explained by a
metal spectrum with low line density. This is for instance the case of
HD~35008 of Table~\ref{Table_Peculiars}, where Balmer line analysis
gives $\Bz = -340 \pm 72$\,G, and metal line analysis gives $\Bz =
-273 \pm 293$\,G. Here note that metal line analysis does not confirm
the magnetic field detection, but is still consistent with the field
measured via Balmer line analysis. This star's
spectrum is not rich in metal lines, explaining the large error
bar of \Bz\ obtained through the metal line method.  By contrast, a
``Dnn'', or a ``dnn'' sequence in a star with a large blocking factor
would prompt a re-analysis of the data reduction, or defer judgment
about field detection to further observations.

\section{Discussion}\label{Sect_Discussion}
\begin{figure}
\includegraphics*[angle=270,width=9.0cm]{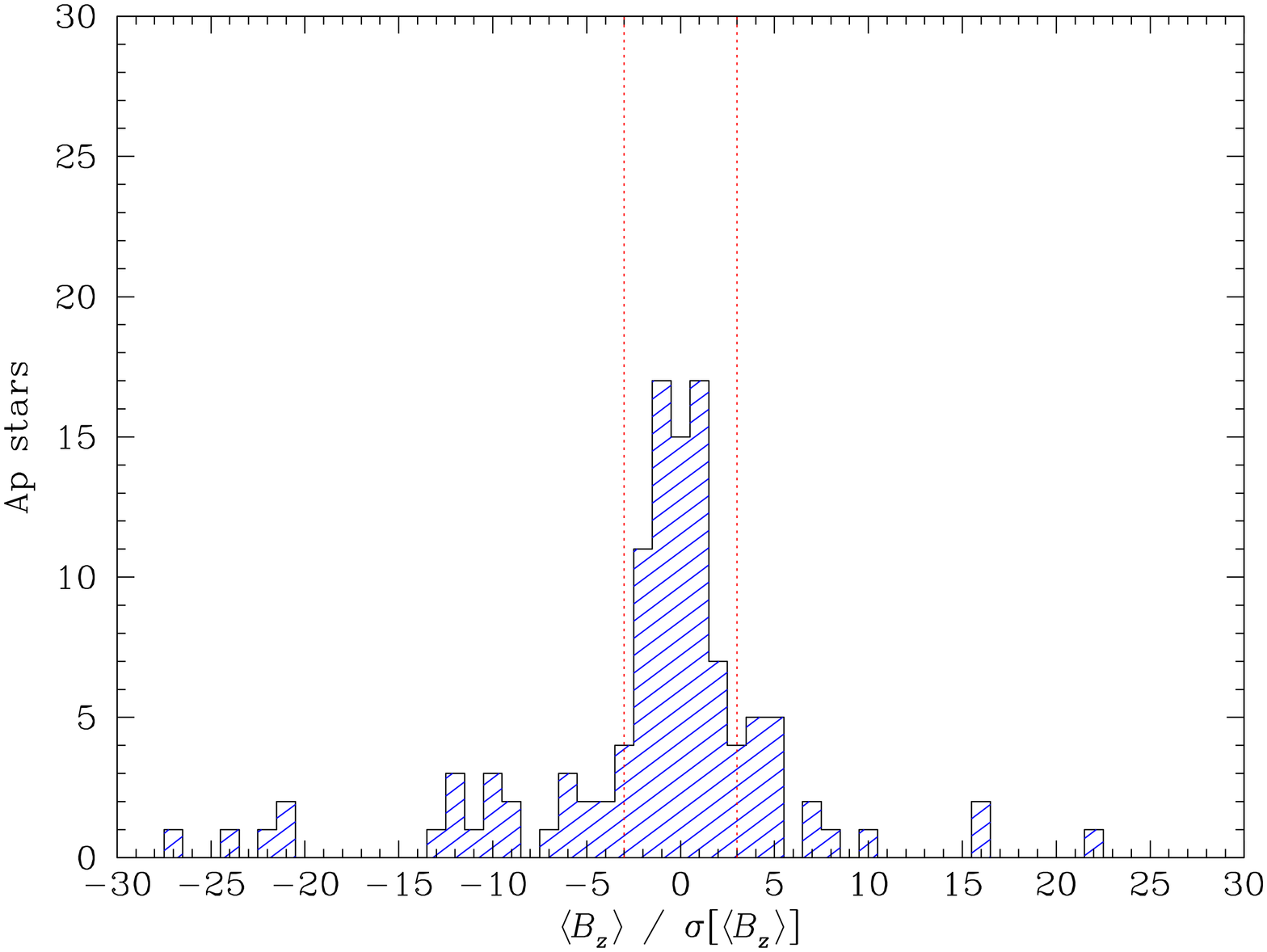}\\
\includegraphics*[angle=270,width=9.0cm]{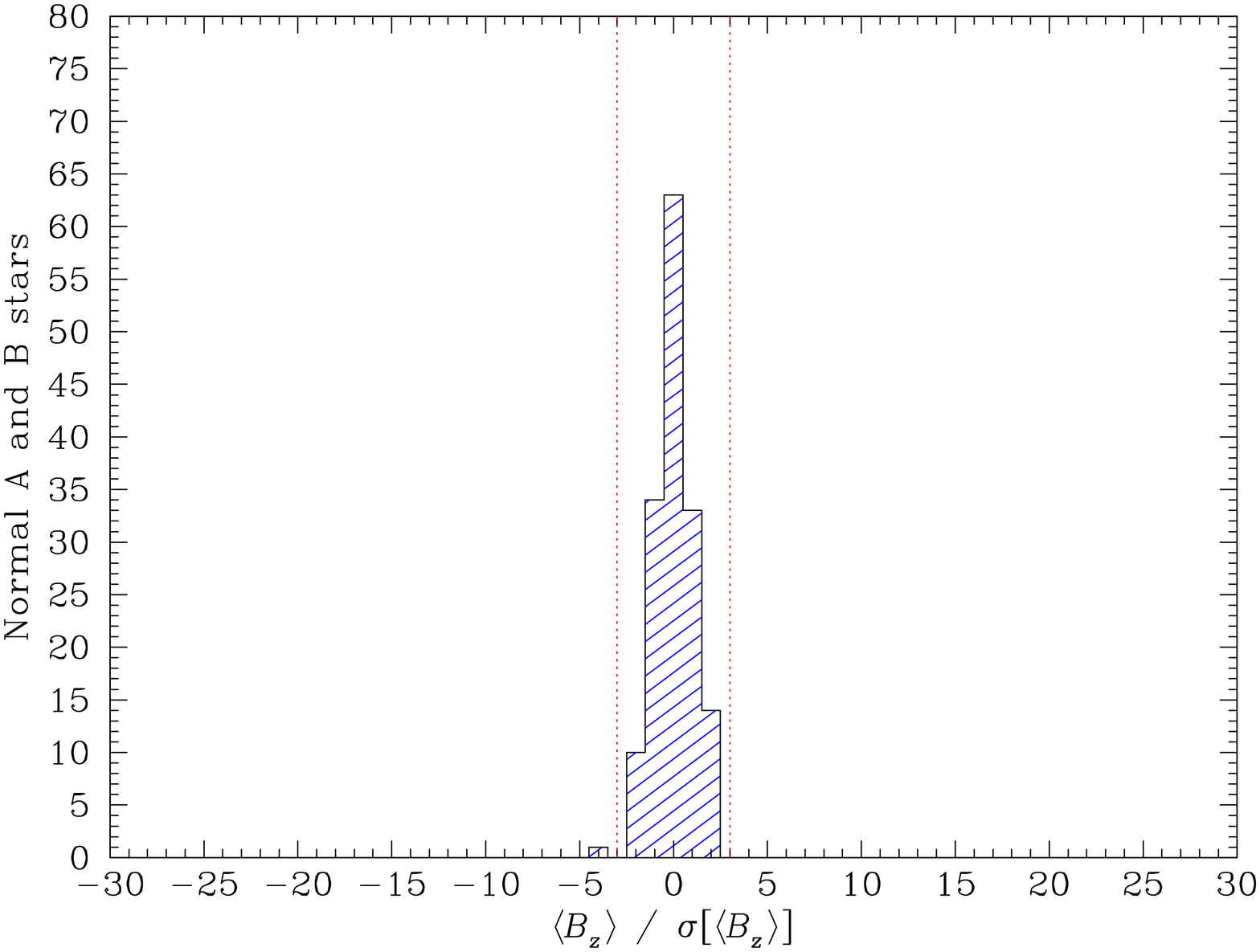}
\caption{\label{FigHistoNorm} Distribution of \Bz\ values normalised
to their error bars measured from Balmer lines in the Ap stars (upper
panel) and in the normal A and B-type stars (lower panel). Five stars
have been omitted from the upper panel (Ap stars) because they are off
scale. The dotted lines correspond to the 3-$\sigma$ detection level.}
\end{figure}

In Fig.~\ref{FigHistoNorm} we show the histograms of the \Bz\ values
calculated via Balmer line analysis and normalized to their error
bars. 
The upper panel refer to the Ap stars, and the lower panel refers to
the normal early-type stars. The top histogram shows that in 58\,\% of
the observed Ap stars we obtained a null detection. However, it would
be incorrect to infer that only $\sim 40$\,\% of the Ap stars are
magnetic. First, targets of Table~\ref{Table_Peculiar_Names} have not
been carefully checked for correct spectral classification, and 
some may not be chemically peculiar. Secondly, and
most importantly, our magnetic field detection threshold is usually
$\sim 250$\,G, whereas observations based on high resolution
spectropolarimetry have shown that a number of Ap stars exhibit a smaller
longitudinal field (e.g., Auri\`{e}re et al.\ \cite{Auretal04}).
Finally, some of our targets may have been observed at rotation
phases at which the 
longitudinal field is small. This appears to be the case for HD~74169, for
which we detected a field in only one of two measurements.  More
detailed statistical considerations will be presented in Paper~II.

\begin{figure*}
\scalebox{0.92}{\includegraphics*[1.0cm,12.5cm][23cm,24.5cm]{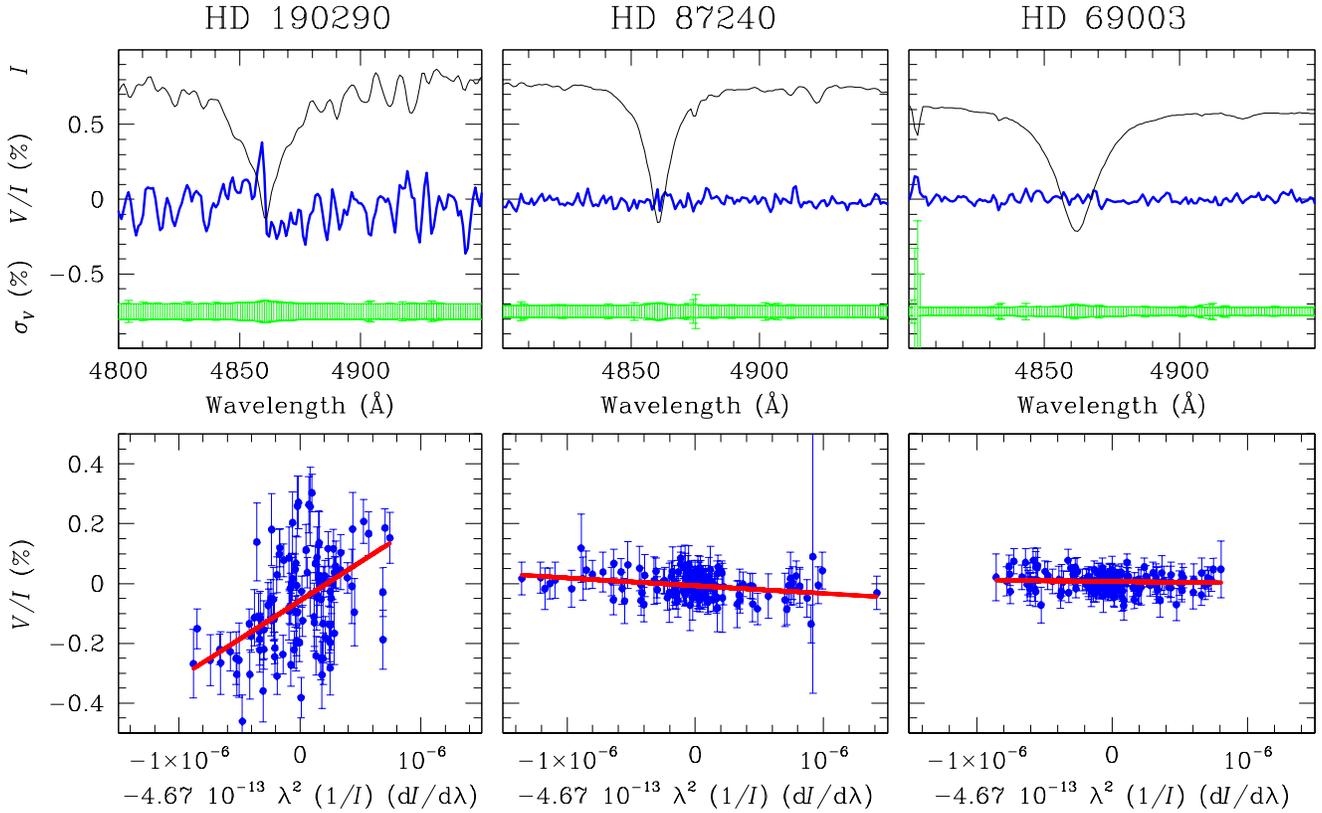}}
\caption{\label{FigSpectra} Top panels: Stokes~$I$ (thin solid lines,
arbitrary units) $V/I$ profiles (thick solid line, in \%), and error
bar associated to $V/I$, for three stars observed during the
survey. Bottom panels: the corresponding best-fit to the Balmer lines
obtained by minimizing the $\chi^2$ of Eq.~(\ref{EqChiSquare}). The
longitudinal field measured in HD~190290 (left panels) is $\Bz = +2429
\pm 110$\,G; the field measured in HD~87240 (middle panels) is $-257
\pm 58$\,G; the field measured in HD~69003 (right panels) is $-48 \pm
53$\,G.}
\end{figure*}

Conversely, the bottom histogram of Fig.~\ref{FigHistoNorm} shows that
no star among the normal A and B-type shows evidence for magnetic
field to a typical $3\,\sigma$ detection limit of 400\,G. This is
fully consistent with the null detections obtained by Landstreet
(\cite{Lan82}) and by Shorlin et al.~(\cite{Shoetal02})
in their surveys based on H$\beta$ photopolarimetry, and high
resolution spectropolarimetry of metal lines, respectively.
Landstreet (\cite{Lan82}) observed 31 targets (including normal A and
B stars, Am, HgMn and $\lambda$ Bootis stars) providing measurements
with a median error of 65\,G. Shorlin et al.\ (\cite{Shoetal02})
observed 64 targets with a typical error between 15 and 50\,G.  Our
survey has a larger detection threshold, but is based on a larger
sample. Therefore, our results bring substantial additional support to
the common belief that Ap stars are the only objects of the middle
main sequence characterized by strong, globally organized, magnetic fields.

Statistical data about rotation velocities show that the typical
\vsini\ values for normal A and B-type stars are in the range
$100-200$\,\kms, depending on the stellar mass. It is therefore
reasonable to assume that our targets have \vsini\ within the same
range. Considering that observations based on high resolution
spectropolarimetry are not suitable for field detection in rapidly
rotating objects (say $\vsini \ga 60$\,\kms), our work represents the
most extensive survey for magnetic fields in normal early-type stars,
and is not biased toward slowly rotating stars. The null result of
this survey suggests that magnetic field organized on large scales are
not common in non-Ap stars. Finally, and most important, the null
detection obtained in non-Ap stars shows strong evidence that the
diagnostic techniques used in this paper are not prone to spurious
detections. The only exception is the star CD-60 1996, which deserves
further investigation.  Note however that with this many stars,
statistically one should expect at least one measurement for which
$\vert \Bz/\sigma[\Bz] \vert > 3$, even in absence of any magnetic fields.

Examples of reduced data and \Bz\ determinations for three selected
stars are illustrated in Fig.~\ref{FigSpectra}, which shows $I$ and
$V/I$ around H$\beta$, as well as the best fit obtained applying
Eq.~(\ref{EqBz}) to all the Balmer lines. HD~190290 (left panels)
exhibits a strong longitudinal field: $\Bz = +2429 \pm
110$\,G. HD~87240 (middle panels) has a magnetic field close to the
detection limit: $\Bz = -257 \pm 58$\,G. HD~69003 (right panels) is a
normal A-type star, and we did not detect a magnetic field ($\Bz = -48 \pm
53$\,G). Note that a simple inspection to the $V/I$
profile of HD~87240 (middle panels) does not show any obvious
polarization signal. The magnetic field is detected by
the linear correlation between $V/I$ and the quantity $1/I\
\times {\rm d}I/{\rm d} \lambda$. 

Table~\ref{Table_NonA} reports a field detection in the star HD~298045
which is of spectral type M3. Although a field detection in a M giant
would be an exciting discovery, this result must be taken very
cautiously. The spectrum of the star is contaminated by a strong
reflection from the LADC (see Sect.~\ref{Sect_Prepro}), and, although
we have measured \Bz\ using only those spectral regions that appear not
contaminated, it is clear that the validity of our measurement is
highly questionable. In any case, it would be certainly interesting to
extend our technique also to late type stars.

\section{Conclusions}\label{Sect_Conclusions}
We have carried out a survey of magnetic fields in A- and B-type stars
belonging to open clusters and associations. The mean longitudinal
magnetic field \Bz\ has been measured in 257 stars, via the analysis
of their circularly polarized spectra obtained with the FORS1
instrument of the ESO VLT.  The results of this survey will be used,
in Paper~II, to study how magnetic fields of early-type stars change
as stars evolve through the main sequence phase.

For the benefit of other FORS1 observers, we have discussed in depth
the technical details of our strategy for observations, data reduction
and data analysis. In particular, we have shown that a combined
analysis of metal and H Balmer lines represents in many cases the best
strategy to detect a weak stellar magnetic field.

Among the observed targets, 97 are candidate Ap cluster member
stars. A magnetic field was detected in 41 Ap stars, 37 of which were
not previously known as magnetic stars. In none of the remaining 160
non-peculiar stars have we detected a field. Among these, 138 are
presumably normal A and B type stars, and 22 are late-type stars. The
fact that no field was detected in any of the stars that are
traditionally considered non magnetic convincingly shows that the
techniques used in this paper are not prone to spurious detections. At
the same time, this work represents one of the largest surveys for
magnetic fields in {\em normal} A and B-type stars. The precision of
this survey is not as high as that obtained through high resolution
spectropolarimetry, but our survey is \textit{not} biased toward
slowly rotating stars.

Before this work, only 13 candidate Ap cluster members, together with
another 25 Ap stars belonging to the Ori B and Sco-Cen association,
had been searched for a magnetic field. Our survey has obtained
observations for an additional 90 Ap possible cluster members, and
added about 30 more clusters to the list of those that have been
searched for magnetic stars.  For the first time we have now enough
observational material to perform a preliminary search for links
between magnetic field and evolution of Ap stars. This will be the
subject of Paper~II.

\begin{acknowledgements}
SB, JDL, VA, and GAW are very grateful to W.W.~Weiss for his kind
hospitality at the Institut f\"{u}r Astronomie, Universit\"{a}t
Wien.  We wish to thank O.\ Kochukhov, F.\ Patat, and Th.~Szeifert
for useful discussions and suggestions. SB acknowledges DGDF 05/02
granted for a science leave in Wien.  JDL and GAW acknowledge
research support from the Natural Sciences and Engineering Research
Council of Canada. JDL and VA acknowledge funding from the
Scientific Visitors Programme of ESO Chile. This research has made
use of the Simbad database, operated at CDS, Strasbourg, and of the
WEBDA database, developed by J.\ Mermilliod and maintained at the
Vienna Observatory by E.\ Paunzen.
\end{acknowledgements}


\Online
\input{4223Appendix.tex}

\input{4223tableA1.tex}
\input{4223tableA2.tex}
\input{4223tableA3.tex}

\input{4223tableA4.tex}

\input{4223tableA5.tex}
\end{document}

%% file: 4223table1.tex
\caption{\label{Table_HD94660} \Bz\ measurements of the $V=6.1$ Ap star
HD~94660 = KQ~Vel. Date, UT, and signal to noise ratio (SNR) are 
calculated as explained in the caption of Table~\ref{Table_Peculiars}. Observations
obtained in P66 had been already published by Bagnulo et al.\ (\cite{Bagetal02}).}
\begin{center}
\begin{tabular}{ll@{\ \ }lr@{\,$\pm$}lr@{\,$\pm$}lr@{\,$\pm$}llc}
\hline\hline
   & &    &\multicolumn{2}{c}{\Bz\ (G)}
          &\multicolumn{2}{c}{\Bz\ (G)}
          &\multicolumn{2}{c}{\Bz\ (G)}
    &      \\ 
\multicolumn{1}{c}{Period}&
\multicolumn{1}{c}{DATE}&
\multicolumn{1}{c}{UT}  &
\multicolumn{2}{c}{Balmer lines} &
\multicolumn{2}{c}{metal lines}  &
\multicolumn{2}{c}{full spectrum}&
\multicolumn{1}{c}{SNR} &  Grism \\
\hline
P66 & 2001-03-22 & 23:54:00 &$-$2085 & 85 &$-$2100 &100 &$-$2260 & 65 & 1550 &600\,B \\
P68 & 2002-02-04 & 08:43:13 &$-$2335 & 57 &$-$2141 & 52 &$-$2226 & 37 & 2100 &600\,B \\
P68 & 2002-02-04 & 08:56:09 &$-$2083 & 75 &$-$2516 & 34 &$-$2439 & 31 & 2150 &600\,R \\
P70 & 2003-02-08 & 09:41:49 &$-$2574 & 57 &$-$2051 & 48 &$-$2260 & 38 & 2150 &600\,B \\
P74 & 2005-01-30 & 09:33:45 &$-$2432 & 50 &$-$2002 & 43 &$-$2190 & 32 & 2300 &600\,B \\
\hline
\end{tabular}
\end{center}

%% file: 4223table2.tex
\caption{\label{Table_Cluster_List} List of open clusters observed in this
survey, with approximate ages (col.~4) and distance moduli $d =
\vert M - m \vert$ (col.~5) extracted from WEBDA. Column~6 gives the number of
candidate peculiar stars and col.~7 the number of non chemically
peculiar A and B stars that we observed in each cluster.}
\begin{tabular}{l@{\ \ }ll@{\ \ }r@{\ \ }rr@{\ \ }r}
\hline \hline
                             &&
      &        &    &    &non\\
\multicolumn{2}{c}{RA \& DEC} &
 NAME &$\log t$&$d$ & Ap & Ap\\
\hline
00 04 07 & $-$29 50 00 &  Blanco\,1      & 7.796   &  7.18    &  1 & 0 \\
04 48 27 & $+$10 56 12 &  NGC\,1662      & 8.625   &  9.14    &  1 & 0 \\
05 26    & $+$03 05    &  Ori\,OB1a ass. & 7.07    &  7.6     &  2 & 0 \\
05 36    & $-$01 12    &  Ori\,OB1b ass. & 6.23    &  8.37    &  4 & 0 \\
05 35    & $-$04 50    &  Ori\,OB1c ass. & 6.66    &  8.52    &  8 & 3 \\
05 27    & $-$05 23    &  Ori\,OB1d ass. & 6.0     &  8.52    &  2 & 1 \\
06 08 24 & $+$13 57 54 &  NGC\,2169      & 7.067   & 10.73    &  1 & 4 \\
06 28 01 & $-$04 50 48 &  NGC\,2232      & 7.727   &  7.87    &  1 & 0 \\
06 31 55 & $+$04 56 30 &  NGC\,2244      & 6.896   & 12.23    &  1 & 3 \\
06 46 01 & $-$20 45 24 &  NGC\,2287      & 8.385   &  9.29    &  3 & 2 \\
06 54 12 & $-$24 25 00 &  Collinder\,121 & 7.054   &  8.45    &  1 & 0 \\
07 02 42 & $-$08 23 00 &  NGC\,2323      & 8.096   & 10.50    &  1 & 2 \\
07 08 06 & $-$10 37 00 &  NGC\,2343      & 7.104   & 10.48    &  0 &28 \\  
07 15 20 & $-$30 41 00 &  Collinder\,132 & 7.080   &  8.48    &  1 & 0 \\
07 36 35 & $-$14 29 00 &  NGC\,2422      & 7.861   &  8.67    &  4 & 2 \\
07 43 12 & $-$38 24 00 &  NGC\,2451a     & 7.78    &  6.41    &  5 & 1 \\
07 56 15 & $-$30 03 48 &  NGC\,2489      & 7.264   & 14.15    &  2 & 1 \\
07 58 04 & $-$60 45 12 &  NGC\,2516      & 8.052   &  8.37    &  7 &31 \\
08 12 15 & $-$37 35 42 &  NGC\,2546      & 7.874   & 10.23    &  5 & 5 \\
08 40 32 & $-$53 02 00 &  IC\,2391       & 7.661   &  6.24    &  2 & 0 \\
08 47 54 & $-$47 27 00 &  Trumpler\,10   & 7.542   &  8.24    &  1 & 0 \\
09 33 11 & $-$53 23 54 &  NGC\,2925      & 7.85    &  9.69    &  1 & 2 \\
10 02 36 & $-$60 07 12 &  NGC\,3114      & 8.093   & 10.01    &  6 & 4 \\
10 21 22 & $-$54 21 22 &  NGC\,3228      & 7.932   &  8.76    &  1 &11 \\
10 37 54 & $-$59 11 00 &  vdb-Hagen 99   & 7.605   &  8.70    &  1 & 1 \\
10 42 58 & $-$64 24 00 &  IC\,2602       & 7.507   &  6.11    &  1 & 0 \\
10 42 04 & $-$59 55 00 &  Collinder\,228 & 6.830   & 12.77    &  2 & 2 \\
11 05 39 & $-$58 45 12 &  NGC\,3532      & 8.492   &  8.55    &  3 & 3 \\
14 07 27 & $-$48 20 36 &  NGC\,5460      & 8.207   &  9.44    &  3 & 2 \\
14 35 37 & $-$56 37 06 &  NGC\,5662      & 7.968   & 10.08    &  2 & 5 \\
16 18 50 & $-$57 56 06 &  NGC\,6087      & 7.976   & 10.29    &  2 & 3 \\
16 35 47 & $-$45 38 36 &  NGC\,6178      & 7.248   & 10.71    &  2 & 0 \\
16 41 20 & $-$48 45 48 &  NGC\,6193      & 6.775   & 11.79    &  2 & 2 \\
17 04 41 & $-$37 59 06 &  NGC\,6281      & 8.497   &  8.86    &  2 & 1 \\
17 34 48 & $-$32 34 00 &  NGC\,6383      & 6.962   & 10.89    &  2 & 6 \\
17 40 20 & $-$32 15 12 &  NGC\,6405      & 7.974   &  8.88    &  4 & 9 \\
17 53 51 & $-$34 47 36 &  NGC\,6475      & 8.475   &  7.71    &  3 & 0 \\
18 27 15 & $+$06 30 30 &  NGC\,6633      & 8.629   &  8.44    &  2 & 0 \\
18 31 47 & $-$19 07 00 &  IC\,4725       & 7.965   & 10.44    &  4 & 4 \\
20 17 19 & $-$79 02 00 &  Mel\,227       & 8.57    &  5.52    &  1 & 0 \\
\hline
\multicolumn{5}{l}{TOTAL}                                     & 97 &138\\
\hline  
\end{tabular}

%% file: 4223table3.tex
\caption{\label{Table_Log} Log of the various observing runs with
FORS1 at the ESO VLT, dedicated to the open cluster survey}
\begin{tabular}{lrllll}
\hline \hline
\multicolumn{1}{c}{Period}         & 
\multicolumn{1}{c}{Program ID}     &  
\multicolumn{1}{c}{Dates}          &
\multicolumn{1}{c}{Observing Mode} & 
\multicolumn{1}{c}{Telescope}      &
\multicolumn{1}{c}{Grism}         \\
\hline
P68 &  68.D-0403  & 2002 February 2/3 and 3/4 & Visitor & UT3 Melipal & 600\,R\\
P70 &  70.D-0352  & 2003 February 7/8 and 8/9 & Visitor & UT1 Antu    & 600\,B 
                                                                     \& 600\,R\\
P72 & 272.D-5026  & 2003 February 28/29 to March 2/3 & Service & UT1 Antu 
                                                                      & 600\,B\\
P73 &  73.D-0498  & 2004 May 28/29 to August 23/24   & Service & UT1 Antu 
                                                            and UT2 Kueyen
                                                                      & 600\,B\\
P74 &  74.D-0488  & 2005 January 28/29 and 29/30     & Visitor & UT2 Kueyen 
                                                                      & 600\,B\\
\hline
\end{tabular}

%% file: 4223Appendix.tex

\appendix
\section{The standard deviation of $\overline{V/I}$}\label{app:sigma_v}

We begin by defining, for brevity and clarity, the following quantities:
\begin{eqnarray}
  v & \equiv & \frac{V}{I} 
  \label{eq:v_def}\; , \\
  \phi^{\pm}  & \equiv & 
  \left(\frac{\fo-\fe}{\fo+\fe}\right)_{\alpha=\pm 45^\circ}
  \label{eq:phi_def}\; . 
\end{eqnarray}
With these definitions, Eq.~(\ref{StokesUno}) becomes:
\begin{equation}
  \label{eq:StokesShort}
  v = \phim - \phip \;.
\end{equation}

We now consider a sequence of $n$ exposures taken with $\alpha=-45^\circ$,
and $m$ exposures taken with $\alpha=+45^\circ$, thus obtaining a set of
$K=n\,m$ pairs, $v_{ij}=\phim_i-\phip_j$. In general we have that $m\neq n$,
if for instance one or more data point have been rejected by the
$\sigma$-clipping algorithm described in Sec.~\ref{SecSigmaClipping}.

From such a set of measurements, we estimate the expected value for $v$
using Eq.~(\ref{EqFinalV}), that is:
\begin{equation}
  \label{eq:vbar_def}
  \bar{v} \equiv 
  \frac{\sum_{ij}v_{ij}/\sigma^2_{ij}}{\sum_{ij} 1/\sigma^2_{ij}}
  \; ,
\end{equation}
where we have shortened the notation for
$\Var{v_{ij}}=\Var{\phim_i}+\Var{\phip_j}$ with $\sigma^2_{ij}$.

We now wish to estimate the variance of the statistical variable
$\bar{v}$:
\begin{eqnarray*}
  \lefteqn{\Var{\bar{v}} = 
    \Mean{\left(\bar{v}-\Mean{v}\right)^2}  = 
    \left(\sum_{i=1}^n\sum_{j=1}^m\sigma^{-2}_{ij}\right)^{-2}\times} \\
  & & 
  \left(\sum_{i,h=1}^n\sum_{j,k=1}^m
    \sigma^{-2}_{ij}\sigma^{-2}_{hk} 
    \Mean{(\phim_i-\phip_j-\Mean{v})(\phim_h-\phip_k-\Mean{v})}
  \right) \; ,
\end{eqnarray*}
where $\Mean{v}$ is of course the mean value of $v$ and of $\bar{v}$.
Since, obviously, $\Mean{v}=\Mean{\phim}-\Mean{\phip}$ (from
Eq.~\ref{eq:StokesShort}), and since the various exposures are statistically
uncorrelated (thus the covariance $\Covar{\phim}{\phip}=0$), it is easy to
derive:
\begin{eqnarray}
  \lefteqn{\Var{\bar{v}}  = 
    \left(\sum_{i=1}^n\sum_{j=1}^m\sigma^{-2}_{ij}\right)^{-2} \times}
  \nonumber \\
  & & 
  \left[\sum_{i=1}^n\sum_{j=1}^m\sigma^{-2}_{ij}
    \left(
      \Var{\phim_i}\sum_{k=1}^m\sigma^{-2}_{ik} + 
      \Var{\phip_j}\sum_{h=1}^n\sigma^{-2}_{hj}
    \right)
  \right] 
  \label{eq:vbar_err} \; .
\end{eqnarray}
If all errors of individual exposures can be assumed equal:
\begin{equation}
  \Var{\phim_i}=\Var{\phip_j}=\sigma^2_{ij}/2=\sigma^2/2 \; ,  
  \label{eq:sigma_equal}
\end{equation}
for $i=1,\ldots n$ and $j=1,\ldots m$, Eqs.~(\ref{eq:vbar_def})
and~(\ref{eq:vbar_err}) become, respectively:
\begin{eqnarray}
  \bar{v} & = & \frac{1}{n\, m} \sum_{ij} v_{ij} %
  = \overline{\phim} - \overline{\phip}
  \label{eq:vbar_def_simple} \\
  \Var{\bar{v}} & = & \frac{\sigma^2}{2} \left(\frac{1}{n} + \frac{1}{m}\right)
  \label{eq:vbar_err_simple} \; .
\end{eqnarray}
The latter equation, in particular, reduces to Eq.~(\ref{EqPhotonNoise}) if
$n=m$. In this case, it is worth noting that at the denominator of
Eq.~(\ref{EqPhotonNoise}) there appears $m$ instead of the number of exposure
pairs $K=m^2$, as it would naively be expected from the analogy with the
average of $m^2$ independent measures.  The reason is obviously that the
$m^2$ combinations are not all completely independent from each other, since
each exposure is counted $m$ times in Eqs.~(\ref{EqFinalV})
and~(\ref{eq:vbar_def}).

It would be desirable to find an alternative to Eq.~(\ref{eq:vbar_err}) that
does not depend explicitly on the knowledge of the statistical errors of the
individual exposures.  We seek an estimator of the form:
\begin{displaymath}
  s^2 = \mathcal{C}\; \sum_{ij} \left( v_{ij} - \bar{v} \right)^2 \; ,
\end{displaymath}
where $\mathcal{C}$ is a constant such that the statistical estimator $s^2$
is not biased, i.~e.\, its mean value coincides with the expected value
\Var{\bar{v}}.  We find such a constant under the assumption that all errors
of individual exposures are assumed equal (Eq.~\ref{eq:sigma_equal}):
\begin{eqnarray*}
  \Mean{s^2} & = & \mathcal{C}\; 
  \Mean{\sum_{ij} \left( v_{ij} - \bar{v} \right)^2} = \\
  & & \mathcal{C}\; 
  \sum_{ij} \Mean{\left[ (v_{ij} - \Mean{v}) - (\bar{v} - \Mean{v}) \right]^2} = \\
  & & \mathcal{C}\; \left[
    n\,m\left(
      \Var{v} + \Var{\bar{v}}
    \right) - 
    2n\,m\,\Mean{(\bar{v} - \Mean{v})^2} 
  \right]= \\
  & & \mathcal{C}\; \left[
    n\,m\left(
      \Var{v} - \Var{\bar{v}}
    \right)
  \right]= \\
  & & \mathcal{C}\; \left[
    m\,(n-1)\,+\,n\,(m-1)
  \right] \sigma^2/2
  \; ,
\end{eqnarray*}
where in the last passage we have used Eq.~\ref{eq:vbar_err_simple}.  From
the above expressions, the estimator $s^2$ is unbiased if it is defined as:
\begin{equation}
  \label{eq:vbar_var_good}
  s^2 = 
  \frac{1}{m\,n\,[2\,n\,m/(n+m)-1]}\; 
  \sum_{ij} \left( v_{ij} - \bar{v} \right)^2
  \; ,
\end{equation}
which, of course, becomes Eq.~(\ref{EqStatistics}) if $m=n$.

%% file: 4223tableA1.tex
\begin{longtable}{lllrlc}
\caption{\label{Table_Peculiar_Names} List of Ap star candidate open
cluster members. Magnitude and spectral types are taken from the
\textit{General catalogue of Ap and Am stars} by Renson et al.
(\cite{Renetal91}), and/or from SIMBAD. In col.~6 we report a flag 
about peculiarity confidence from the same catalogue. Symbol * means `well
knonw Ap star', symbol '?' means `doubtful nature'; we indicate with
`--' the cases in which the star is not included in the Renson et al's
catalogue (in which case spectral type is taken from SIMBAD or
specific literature). The star's actual membership and peculiarity
will be discussed in Paper~II. } \\
\hline\hline
\multicolumn{1}{c}{Cluster} & 
\multicolumn{1}{c}{ID1}     & 
\multicolumn{1}{c}{ID2}     &
\multicolumn{1}{c}{$V$}     &
\multicolumn{1}{c}{Spectral Type} &
\multicolumn{1}{c}{CP prob.} \\
\hline
\endfirsthead
\caption{continued.} \\
\multicolumn{1}{c}{Cluster} & 
\multicolumn{1}{c}{ID1}     & 
\multicolumn{1}{c}{ID2}     &
\multicolumn{1}{c}{$V$}     &
\multicolumn{1}{c}{Spectral Type} &
\multicolumn{1}{c}{CP prob.} \\
\hline
\endhead
\hline
\endfoot
Blanco 1        &HD 225264        &CD-30 19818      & 8.3 & A1 Si\,Sr   &? \\[2mm]
NGC 1966        &HD 30598         &BD+10 649        & 9.1 & A1 Sr\,Cr   &  \\[2mm]
Ori OB1 Ass.    &HD 35008         &BD-01 872        & 7.1 & B9 Si       &? \\      
                &HD 36046         &BD-00 964        & 8.1 & B8 He-weak  &  \\ 	    
                &HD 36540         &BD-04 1162       & 8.1 & B7 He-weak  &  \\	    
                &HD 36549         &BD+01 1053       & 8.6 & B7 He-weak  &  \\  
                &HD 36629         &BD-04 1164       & 7.7 & B3 He-weak  &? \\ 	
                &HD 36918         &BD-06 1231       & 8.3 & B9 He-weak  &? \\  
                &HD 36916         &V1045 Ori        & 6.7 & B8 He-weak Si&*\\  
                &HD 36960         &BD-06 1234       & 4.8 & B0 Si       &  \\	
                &HD 36982         &VM LP Ori        & 8.4 & B2 He       &? \\	
                &HD 290665        &BD-00 1008       & 9.4 & A0 Cr\,Eu\,Sr& \\	
                &HD 37022         &tet01 Ori C      & 5.1 & O6pe        &-- \\	
                &HD 37058         &V359 Ori         & 7.3 & B3 He-weak Sr&*\\	
                &HD 37210         &V1133 Ori        & 8.1 & B8 He-weak Si& \\	
                &HD 37333         &BD-02 1319       & 8.5 & A0          &? \\	
                &HD 37470         &BD-06 1274       & 8.2 & B8 Si       &  \\	
                &HD 37633         &V1147 Ori        & 9.0 & B9 Si\,Eu   &  \\[2mm] 
NGC 2169        &V1356 Ori        &NGC 2169 12      &10.8 & A0 Si       &  \\[2mm]
NGC 2232        &HD 45583         &V682 Mon         & 8.0 & B9 Si       &  \\[2mm]
NGC 2244        &NGC 2244 334     &GSC 00154-02164  &12.9 & B3          &-- \\[2mm]
NGC 2287        &HD 49023         &BD-20 1543       & 8.4 & B9 Hg\,Si   &  \\
                &CPD-20 1640      &NGC 2287 AR 143  &11.1 & A5 Si\,Sr   &-- \\
                &HD 49299         &BD-20 1573       &10.2 & A0 Si\,Sr\,Cr& \\[2mm]
Collinder 121   &HD 51088         &CD-24 4586       & 8.3 & B8 Si       &? \\[2mm]
NGC 2323        &HD 52965         &BD-08 1701       & 9.2 & B8 Si       &? \\[2mm]
Collinder 132   &HD 56343         &CD-31 4243       & 9.2 & B9          &-- \\[2mm]
NGC 2422        &BD-14 2015       &NGC 2422 70      & 9.5 & B9 Sr       &  \\
                &BD-14 2028       &NGC 2422 88      &10.4 & A1 Sr\,Cr   &? \\
                &HD 61045         &BD-14 2029       & 8.0 & B8 Si       &  \\
                &BD-14 2040       &NGC 2422 103     &10.6 & A1 Sr\,Cr   &? \\[2mm]
NGC 2451        &HD 62376         &CD-38 3564       & 6.5 & B7          &-- \\
                &CD-37 3845       &NGC 2451 36      & 8.6 & A0 Si       &? \\
                &HD 62992         &CD-37 3860       & 7.9 & A4 Sr\,Eu   &  \\
                &HD 63079         &CD-37 3868       & 7.0 & B9 Si       &? \\
                &HD 63401         &OX Pup           & 6.3 & B9 Si       &  \\[2mm]
NGC 2489        &NGC 2489 58      &NGC 2489 PM 39   &12.8 & A0          &-- \\
                &NGC 2489 40      &NGC 2489 PM 54   &12.6 & B5          &-- \\
NGC 2516        &HD 65712         &CD-60 1925       & 9.4 & A0 Si\,Cr   &  \\
                &CPD-60944A       &NGC 2516 DAC 73  & 8.3 & B9 Si       &  \\
                &CPD-60944B       &NGC 2516 DAC 74  & 8.8 & B9 Si       &? \\
                &CPD-60 978       &V 391 Car        & 8.9 & A0 Si\,Cr\,Eu& \\
                &HD 65987         &V356 Car         & 7.6 & B9 Si\,Sr   &* \\
                &HD 66295         &V422 Car         & 9.1 & B9 Si       &  \\
                &HD 66318         &NGC 2516 DAC 95  & 9.6 & A0 Eu\,Cr\,Sr&*\\[2mm]
NGC 2546        &NGC 2546 258     &--               &10.6 & --          &-- \\
                &NGC 2546 201     &--               &?    & --          &-- \\
                &CPD-37 1989      &NGC 2546 197     &11.0 & A2          &? \\
                &HD 69004         &CD-37 4414       & 9.4 & B8 Si       &? \\
                &HD 69067         &CD-37 4420       & 8.3 & B8 Si       &  \\[2mm]
IC 2391         &HD 74169         &KR Vel           & 7.2 & A1 Si\,Cr\,Sr&* \\
                &HD 74535         &KT Vel           & 5.6 & B9 Si       &* \\[2mm]
Trumpler 10     &HD 75239         &CD-41 4498       & 9.2 & B9 Si       &  \\[2mm]
NGC 2925        &HD 83002         &CD-52 3172       & 9.1 & B8 Si       &  \\[2mm] 
NGC 3114        &HD 87241         &CD-59 2746       & 7.8 & B9 Si       &? \\
                &HD 87240         &CPD-59 1673      & 9.6 & B9 Si       &  \\
                &HD 87266         &CD-59 2755       & 8.2 & B3 Si N     &  \\
                &HD 304841        &CPD-59 1717      &10.0 & A           &? \\
                &HD 304842        &V424 Car         & 9.7 & B8 Si       &  \\
                &HD 87405         &CD-59 2803       & 8.5 & B9 Si       &  \\[2mm]
NGC 3228        &HD 89856         &CD-51 4685       & 9.1 & B8          &? \\[2mm]
vdB-Hagen88     &HD 92190         &CD-58 3407       & 8.6 & B8          &-- \\[2mm]
IC 2602         &HD 92385         &V 407 Car        & 6.7 & B9 Si       &  \\[2mm] 
Collinder 228   &Collinder 228 30 &Collinder 228 CP4&10.8 & B1          &-- \\     
                &HD 305451        &CPD-59 2496      &10.5 & B9 Si       &? \\[2mm]
NGC 3532        &HD 96040         &CPD-57 4168      &10.7 & A           &? \\
                &HD 96729         &CPD-58 3157      &10.0 & B9 Si       &  \\
                &HD 303821        &NGC 3532 241     &11.7 & A           &? \\[2mm]
NGC 5460        &HD 122983        &CD-47 8870       & 9.9 & A0          &? \\
                &HD 123183        &CD-47 8895       & 9.9 & A0          &-- \\
                &HD 123225        &CD-47 8901       & 8.9 & B9 Si       &? \\[2mm]
NGC 5662        &CPD-56 6330      &NGC 5662 85      &10.6 & A2          &  \\
                &HD 127866        &CD-56 5516       & 8.3 & B8          &-- \\[2mm]
NGC 6087        &CPD-57 7817      &NGC 6087 25      &10.0 & --          &? \\
                &HD 146555        &CPD-57 7872      &10.3 & B9 Si       &  \\[2mm]
NGC 6178        &HD 149257        &CD-45 10768      & 8.5 & B5 He       &  \\
                &HD 149277        &CD-45 10769      & 8.4 & B2          &-- \\[2mm]
NGC 6193        &CD-48 11051      &CPD-48 8684      &10.4 & B1          &-- \\
                &CD-48 11059      &NGC 6193 VF 33   &10.7 & B3          &-- \\[2mm]
NGC 6281        &HD 322676        &CD-37 11203      &10.2 & A0          &-- \\
                &HD 153948        &V948 Sco         & 9.3 & A2 Si       &  \\[2mm]
NGC 6383        &NGC 6383 26      &NGC 6383 FJL 5   &12.9 & A           &-- \\
                &HD 317857        &NGC 6383 3       &10.3 & A2          &-- \\[2mm]
NGC 6405        &HD 318107        &V970 Sco         & 9.3 & B8          &? \\
                &HD 318100        &V971 Sco         & 9.8 & B9          &  \\
                &CD-32 13119      &NGC 6405 7       &10.9 & --          &? \\
                &HD 318095        &--               &10.9 & A0          &? \\[2mm]
NGC 6475        &HD 162305        &CD-34 12154      & 7.8 & B9          &? \\
                &HD 320764        &CD-34 12161      & 8.9 & A6          &? \\
                &HD 162725        &V951 Sco         & 6.4 & B9 Si\,Cr   &* \\[2mm]
NGC 6633        &HD 169959        &BD+06 3762       & 7.8 & A0 Si       &  \\
                &HD 170054        &BD+06 3772       & 8.2 & B7 Si       &  \\[2mm]
IC 4725         &BD-19 5044L      &CPD-19 6897      &10.2 & B8          &? \\
                &BD-19 5046       &CPD-19 6905      & 9.0 & A1          &-- \\
                &HD 170836        &BD-19 5052       & 9.0 & B7          &? \\
                &HD 170860A       &BD-19 5058       & 9.4 & B8          &-- \\[2mm]
Melotte 227     &HD 190290        &CK Oct           & 9.9 & A0 Eu\,Sr   &  \\
\hline
\end{longtable}

%% file: 4223tableA2.tex
\begin{longtable}{lllrl}
\caption{\label{Table_Normal_Names} List of the observed normal cluster A and B-type
stars. Actual membership has not been checked}\\
\hline\hline
\multicolumn{1}{c}{Cluster} & 
\multicolumn{1}{c}{ID1}     & 
\multicolumn{1}{c}{ID2}     &
\multicolumn{1}{c}{$V$}     &
\multicolumn{1}{c}{Spectral Type}\\
\hline
\endfirsthead
\caption{continued.}\\
\hline\hline
\multicolumn{1}{c}{Cluster} & 
\multicolumn{1}{c}{ID1}     & 
\multicolumn{1}{c}{ID2}     &
\multicolumn{1}{c}{$V$}     &
\multicolumn{1}{c}{Spectral Type}\\
\hline
\endhead
\hline
\endfoot
Ori OB1 Ass.     &HD 36559         &BD-04 1163        & 8.8   &    B9    \\
                 &HD 36671         &BD-04 1165        & 8.7   &    B9    \\
                 &HD 37041         &tet02 Ori A       & 5.1   &   O9.5Vpe\\
                 &HD 37428         &BD-06 1271        & 8.8   &    A0    \\[2mm]
NGC 2169         &HD 252214        &BD+13 1120        & 8.1   &    B2.5V \\
                 &HD 41909         &BD+14 1160        & 8.4   &    B5    \\
                 &HD 252248        &BD+13 1123        & 8.8   &    B2V   \\
                 &HD 252266        &BD+13 1124E       & 9.2   &    B3V   \\[2mm]
NGC 2244         &NGC 2244 336     &GSC 00154-01288   &12.6   &    A0    \\
                 &CPD-20 1637      &NGC 2287 54       &10.9   &    A1V   \\
                 &BD-20 1571       &NGC 2287 55       &10.8   &    A3    \\[2mm]
NGC 2287         &NGC 2287 AR 157  &--                &  ?    &     ?    \\
                 &CPD-20 1645      &NGC 2287 AR 159   &  ?    &     ?    \\[2mm]
NGC 2323         &HD 52980         &BD-08 1796        & 8.4   &    B9    \\
                 &BD-08 1708       &NGC 2323 51       & 9.9   &    B6    \\[2mm]
NGC 2343         &CSI-10-07049     &NGC 2343 13       &10.9   &    --    \\
                 &BD-10 1875       &NGC 2343 9        &10.6   &    [A0]  \\
                 &NGC 2343 22      &CSI-10-07051      & 11.8  &     [A0V]\\
                 &NGC 2343 40      &CSI-10-07053 3    & 13.1  &    [F1?V]\\
                 &CSI-10-07053 5   &NGC 2343 35       &12.9   &    [A6V] \\
                 &BD-10 1878       &NGC 2343 12       &10.8   &    [B8/9]\\
                 &NGC 2343 25      &CSI-10-07054 1    & 11.7  &   [A3/4V]\\
                 &NGC 2343 16      &CSI-10-07055 1    & 11.4  &     [A8?]\\
                 &BD-10 1879a      &NGC 2343 7        &10.3   &     [A3V]\\
                 &NGC 2343 34      &CSI-10-07056 5    & 12.7  &     [A6V]\\
                 &NGC 2343 43      &CSI-10-07056 6    & 13.6  &     [F?] \\
                 &NGC 2343 28      &CSI-10-07055 2    & 12.5  &     [A5V]\\
                 &NGC 2343 37      &CSI-10-07055 3    & 13.1  &     [A8?]\\
                 &HD 54304         &BD-10 1882        &  9.9  &      A   \\
                 &NGC 2343 27      &CSI-10-07056 4    & 12.4  &   [A5/6V]\\
                 &NGC 2343 26      &CSI-10-07057 4    & 12.3  &      [A7]\\
                 &NGC 2343 6       &CSI-10-07057 1    &10.3   &     [B8?]\\
                 &NGC 2343 36      &CSI-10-07058 8    & 13.0  &   [A7/8?]\\
                 &NGC 2343 17      &CSI-10-07057 3    & 11.7  &   [A2/3V]\\
                 &NGC 2343 10      &CSI-10-07057 2    & 10.5  &   [B8/9?]\\
                 &NGC 2343 31      &CSI-10-07058 7    & 12.6  &    [A6]  \\
                 &BD-10 1883B      &--                &       &          \\
                 &BD-10 1883A      &NGC 2343 5        & 10.1  &    [B9]  \\
                 &NGC 2343 18      &CSI-10-07058 1    & 11.4  &     [A4V]\\
                 &NGC 2343 38      &CSI-10-07058 6    & 13.0  &    [A7/8]\\
                 &HD 54360         &BD-10 1884        &  9.4  &    A0    \\
                 &BD-10 1885B      &NGC 2343 11       &10.7   &    B9/A0?\\
                 &HD 54388         &BD-10 1887        & 8.4   &    A3    \\[2mm]
NGC 2422         &HD 60940         &BD-14 2012        & 8.7   &  B7/B8III\\
                 &HD 60996         &BD-14 2019        & 8.7   &    BV    \\[2mm]
NGC 2451         &HD 62974         &CD-37 3855        & 8.3   &     A3V  \\[2mm]
NGC 2489         &NGC 2489 59      &GSC 07119-01134   &12.9   &    B9    \\[2mm]
NGC 2516         &HD 65691         &NGC 2516 DAC 9    &  9.0  &    B8/B9V\\
                 &CPD-60 942       &NGC 2516 DAC 10   &10.1   &    A1V   \\
                 &CD-60 1929       &NGC 2516 DAC 12   & 8.5   &    B9III \\
                 &CD-60 1932       &NGC 2516 DAC 14   & 9.9   &    A0V   \\
                 &HD 65869         &CD-60 1937        &  7.7  &    B9V   \\
                 &HD 65896         &NGC 2516 DAC 18   &  9.4  &    A0V   \\
                 &HD 65950         &NGC 2516 DAC 27   &  6.9  &    B8III \\
                 &V373 Car         &NGC 2516 DAC 28   & 9.0   &    B     \\
                 &HD 65949         &NGC 2516 DAC 26   &  8.4  &    B8/B9 \\
                 &BD-60 969        &NGC 2516 DAC 29   & 8.6   &    B9.5IV\\
                 &BD-60 975        &NGC 2516 DAC 504  & 8.9   &    Avar  \\
                 &CPD-60 977       &NGC 2516 DAC 211  &10.4   &    F0V   \\
                 &V410 Car         &NGC 2516 DAC 81   &10.7   &    A7V   \\
                 &V392 Car         &NGC 2516 DAC 37   & 9.5   &    A2V   \\
                 &CD-60 1967       &NGC 2516 DAC 38   & 7.2   &    B9.5IV\\
                 &CD-60 1971       &NGC 2516 DAC 42   & 8.1   &    B8.5V \\
                 &V417 Car         &NGC 2516 DAC 40   &10.4   &    [A6V] \\
                 &CPD-60 984       &NGC 2516 DAC 41   & 9.7   &    A2V   \\
                 &CPD-60 986       &NGC 2516 DAC 901  &10.1   &     [A2V]\\
                 &V418 Car         &NGC 2516 DAC 43   &11.0   &   [A6/7V]\\
                 &CD-60 1974       &NGC 2516 DAC 607  & 9.4   &    A1V   \\
                 &CD-60 1975       &NGC 2516 SBL 753  & 8.4   &    B9V   \\
                 &CD-60 1976       &NGC 2516 DAC 48   & 9.7   &    A0V   \\
                 &CD-60 1979       &NGC 2516 DAC 615  &10.8   &    A3V   \\
                 &CD-60 1978       &NGC 2516 DAC 49   & 8.8   &  B8.5IV-V\\
                 &CD-60 1981       &NGC 2516 DAC 608  &10.7   &    A1Vm  \\
                 &V420 Car         &NGC 2516 DAC 51   &10.4   &    A3V   \\
                 &HD 66137         &NGC 2516 DAC 53   & 7.8   &    B9V:  \\
                 &HD 66194         &V374 Car          & 5.8   &   B2IVnpe\\
                 &CD-60 1996       &NGC 2516 DAC 806  &10.8   &    A8V   \\
                 &CD-60 1999       &NGC 2516 DAC 60   &10.0   &    A2V   \\[2mm]
NGC 2546         &$[$N$75]$ 195    &GSC 07133-00458   &10.5   &    B3:   \\
                 &$[$N$75]$ 196    &--                &10.7   &    B3:   \\
                 &NGC 2546 272     &GEN \#2.25460272  &10.7   &    B3    \\
                 &CD-37 4353       &CPD-37 1983       &  ?    &     ?    \\
                 &HD 69003         &CD-37 4413        & 8.8   &    A0    \\[2mm]
NGC 2925         &HD 298537        &CPD-52 2517       &11.0   &    A     \\
                 &HD 298536        &CPD-52 2522       & --    &    AO    \\[2mm]
NGC 3114         &NGC 3114 233     &--                & --    &    --    \\
                 &CPD-59 1698      &NGC 3114 54       &10.8   &    --    \\
                 &CPD-59 1700      &NGC 3114 56       &11.1   &    --    \\
                 &CPD-59 1703      &NGC 3114 65       &11.0   &    --    \\[2mm]
NGC 3228         &HD 298051        &CD-51 4686        &10.2   &    A1V   \\
                 &HD 89901         &V343 Vel          & 8.4   &  B8/B9III\\
                 &HD 89900         &CD-51 4691        & 8.2   & A0/A1IV/V\\
                 &HD 89915         &CD-51 4693        & 7.8   &    B9.5V \\
                 &HD 89922         &CD-51 4696        & 9.3   &    A4IV/V\\
                 &HD 298047        &CD-51 4695        & 9.2   &    B9V   \\
                 &HD 89938         &CD-51 4698        & 9.3   &    A     \\
                 &HD 89937         &CD-51 4699        & 8.9   &    B6/B7 \\
                 &HD 89956         &CD-51 4701        & 8.2   &    B9IV/V\\
                 &CPD-51 3249      &NGC 3228 15       &11.3   &    [A8V] \\
                 &HD 298053        &CD-51 4702        &10.6   &    A3m   \\[2mm]
vdB-Hagen 88     &HD 303107        &HD 92234          & 9.6   &    B9    \\[2mm]
Collinder 228    &HD 305535        &Collinder 228 25  & 9.4   &    B2.5Vn\\
                 &HD 305543        &DW Car            &10.0   &    B5Iab \\[2mm]
NGC 3532         &HD 96653         &CD-58 3835        & 8.4   &    A0III \\
                 &NGC 3532 447     &--                & 9.7   &    [A5]  \\
                 &HD 96790         &CD-57 3718        &10.1   &    B     \\[2mm]
NGC 5460         &CD-47 8868       &NGC 5460 36       &10.7   &    A0    \\
                 &HD 123201B       &CD-47 8899B       & 9.1   &    A0    \\[2mm]
NGC 5662         &HD 127835        &CD-55 5723        & 9.4   &    B8V   \\
                 &CPD-56 6334      &NGC 5662 104      & 9.9   &    B9V   \\
                 &NGC 5662 126     &CPD-56 6337       &10.7   &     A1V  \\
                 &HD 127900        &CD-55 5726        & 8.8   &  B8II/III\\
                 &HD 127924        &CD-55 5727        & 9.2   &  B8III/IV\\[2mm]
NGC 6087         &NGC 6087 129     &GEN\# +2.60870129 &11.5   &    --    \\
                 &TYC 8719- 717-1  &GSC 08719-00717   &10.2   &    [A0]  \\
                 &HD 146484        &CPC 20 5173       & 9.5   &    A0    \\[2mm]
NGC 6193         &CD-48 11050      &CPD-48 8680       &10.3   &    A2    \\
                 &CD-48 11060      &CPD-48 8694       &10.7   &    B3V   \\[2mm]
NGC 6281         &HD 323673        &CD-37 11212       &10.2   &    A5    \\[2mm]
NGC 6383         &NGC 6383 28      &GEN\# +2.63830028 &12.5   &    A     \\
                 &NGC 6383 700     &GEN\# +2.63830700 &12.7   &    --    \\
                 &NGC 6383 87      &--                &13.7   &    --    \\
                 &HD 317846        &CPD-32 4611       & 9.9   &    B5    \\
                 &NGC 6383 102     &GEN\# +2.63830102 &14.6   &    --    \\
                 &HD 317852        &NGC 6383 21       &11.9   &    A0    \\[2mm]
NGC 6405         &HD 318108        &CD-32 13080       & 9.7   &    B9    \\
                 &HD 318109        &--                & 9.9   &    A0    \\
                 &CD-32 13089      &NGC 6405 53       & 9.9   &    [A4]  \\
                 &CD-32 13093      &NGC 6405 52       &10.2   &    [A0]  \\
                 &V976 Sco         &NGC 6405 31       &11.5   &     [A?] \\
                 &HD 318099        &CD-32 13111       & 9.9   &    A0    \\
                 &HD 320765        &CD-34 12156       & 8.8   &    A2    \\
                 &HD 162678        &CD-34 12219       & 6.4   &    B9V   \\
                 &HD 162724        &V906 Sco          & 6.0   &    B9V+  \\[2mm]
IC 4725          &BD-19 5044 F     &?                 & 9.7   &    B8V   \\
                 &BD-19 5045 ?     &CPD-19 6899       & 9.1   &    B5    \\
                 &BD-19 5044 M     &CPD-19 6901       &10.2   &    B8V   \\
                 &HD 170835        &BD-19 5055        & 8.8   &    B5IV  \\
\hline
\end{longtable}

%% file: 4223tableA3.tex
\begin{longtable}{lllr@{\,$\pm$}lr@{\,$\pm$}lr@{\,$\pm$}llll}
\caption{\label{Table_Peculiars} \Bz\ measurements for chemically
peculiar stars. Columns~2 and 3 give the date and UT at mid-exposure of the
observation, respectively. Column 4, 5, and 6 report the \Bz\
measurement from the Balmer lines, the metal lines, and the full
spectrum, as explained in the text. Column~7 reports the flag for the
\Bz\ detection obtained from Balmer, metal lines, and full
spectrum. The meaning of `D', `d', and `n' is explained in the
text, in Sec.~7. Column 8 reports the signal to noise ratio (SNR) per \AA\
calculated, for stars observed with GRIS\_600\,B, in the wavelength
interval 4975-5025\,\AA\ (close to the red wing of H$\beta$).  For
stars observed with grism 600\,R, the SNR was calculated in the
wavelength interval 6640--6690\,\AA\ (i.e., close to the red wing of
H$\alpha$). Column~9 is used to report flags and notes. Unless
otherwise marked, the star was observed with grism 600\,B}\\
\hline\hline
    &
    &     &\multicolumn{2}{c}{\Bz\ (G)}
          &\multicolumn{2}{c}{\Bz\ (G)}
          &\multicolumn{2}{c}{\Bz\ (G)}
          &    &     &      \\ 
\multicolumn{1}{c}{ID1} & 
\multicolumn{1}{c}{DATE}&
\multicolumn{1}{c}{UT\ \ \ \ \ \ } &
\multicolumn{2}{c}{Balmer lines} &
\multicolumn{2}{c}{metal lines}  &
\multicolumn{2}{c}{full spectrum}&
Flag& 
\multicolumn{1}{c}{SNR} & 
Note \\ 
\hline
\endfirsthead
\caption{continued.}\\
\hline\hline
    &
    &     &\multicolumn{2}{c}{\Bz\ (G)}
          &\multicolumn{2}{c}{\Bz\ (G)}
          &\multicolumn{2}{c}{\Bz\ (G)}
          &    &     &      \\ 
\multicolumn{1}{c}{ID1} & 
\multicolumn{1}{c}{DATE}&
\multicolumn{1}{c}{UT\ \ \ \ \ \ } &
\multicolumn{2}{c}{Balmer lines} &
\multicolumn{2}{c}{metal lines}  &
\multicolumn{2}{c}{full spectrum}&
Flag& 
\multicolumn{1}{c}{SNR} & 
Note \\ 
\hline
\endhead
\hline
\endfoot
HD 225264        &2004-07-31&09:49:40 & $-$48  &  38 &    71  &  72 & $-$23  &  32 & nnn &3900 &     \\
HD 30598         &2004-09-01&07:42:15 &   124  & 120 & $-$55  & 143 &    49  &  90 & nnn &1600 &     \\
                 &2005-01-29&01:54:58 &   121  &  65 &    89  &  94 &   118  &  52 & nnd &2500 &     \\
HD 35008         &2005-01-29&00:22:02 &$-$340  &  72 &$-$273  & 293 &$-$339  &  69 & DnD &2200 &1,M (new)  \\
HD 36046         &2005-01-30&01:21:07 & $-$27  &  83 &$-$110  & 175 & $-$39  &  74 & nnn &2000 &     \\
HD 36540         &2003-02-08&01:40:07 &   301  &  67 &   187  & 115 &   277  &  56 & dnD &2700 &1,M (new)  \\
HD 36549         &2005-01-30&02:01:45 & $-$57  &  80 &$-$206  & 223 & $-$57  &  74 & nnn &1800 &     \\
HD 36629         &2003-02-08&02:08:23 &    87  &  66 & $-$58  &  73 &    15  &  52 & nnn &2600 &    2\\
HD 36918         &2003-02-09&02:10:05 & $-$13  &  58 &    31  & 150 &     6  &  54 & nnn &2700 &    \\
HD 36916         &2003-02-09&01:00:16 &$-$539  &  56 &$-$261  & 123 &$-$488  &  51 & DnD &2700 & 3,M \\
HD 36960         &2003-02-09&01:42:20 & $-$90  &  63 &$-$132  &  64 &$-$115  &  45 & nnn &2900 &     \\
HD 36982         &2003-02-08&01:11:00 &   238  & 239 &$-$192  & 271 &    91  & 193 & nnn & 750 &     \\
HD 290665        &2005-01-29&01:20:42 &$-$1778 &  67 &$-$1588 &  64 &$-$1664 &  44 & DDD &2250 &  M (new)  \\
HD 37022         &2003-02-08&01:11:00 &    57  &  82 &   115  &  82 &    86  &  49 & nnn &4150 &     \\
HD 37058         &2003-02-09&00:41:09 &$-$975  &  77 &$-$664  &  87 &$-$812  &  59 & DDD &2100 & 4,M \\
HD 37210         &2003-02-09&01:21:46 &    29  &  66 &$-$157  & 117 & $-$30  &  57 & nnn &2250 & 1   \\
HD 37333         &2005-01-29&00:44:54 &    23  &  63 &$-$261  & 117 & $-$46  &  55 & nnn &2600 &     \\
HD 37470         &2003-02-08&02:49:06 & $-$62  &  66 &$-$185  & 166 & $-$62  &  60 & nnn &2450 & 1   \\
HD 37633         &2005-01-30&01:41:35 &   295  &  83 &   292  &  97 &   283  &  62 & DdD &2000 &  M (new)  \\
V1356 Ori        &2003-02-08&03:40:54 &$-$3408 & 144 &$-$1290 & 180 &$-$2449 & 113 & DDD &1250 &  M (new)  \\
HD 45583         &2003-09-09&05:40:42 &$-$1518 &  71 &$-$1370 &  58 &$-$1433 &  44 & DDD &2500 &  M (new)  \\
NGC 2244 334     &2003-02-09&03:52:60 &$-$9515 & 196 &$-$4815 & 136 &$-$6215 & 112 & DDD &1100 & 5,M (new)  \\
HD 49023         &2005-01-30&02:21:14 &    79  &  70 & $-$48  & 244 &   100  &  68 & nnn &2000 &     \\
CPD-20 1640      &2005-01-30&03:35:37 &   204  &  92 &    40  & 101 &   157  &  63 & dnn &1850 &     \\
HD 49299         &2004-09-28&08:26:26 &$-$483  &  42 &$-$508  &  37 &$-$659  &  83 & DDD &3300 &  M (new)  \\
                 &2005-01-30&02:50:57 &$-$2720 &  60 &$-$2311 &  77 &$-$2598 &  45 & DDD &2200 &  M  \\
HD 51088         &2004-09-21&09:13:05 &$-$118  &  51 &   103  & 103 & $-$76  &  45 & nnn &2900 &     \\
                 &2005-01-30&05:22:22 &  $-$5  &  74 &   384  & 135 &   102  &  64 & ndd &1900 &     \\
HD 52965         &2005-01-30&05:03:48 & $-$69  &  92 &$-$105  & 217 & $-$64  &  84 & nnn &1800 &     \\
HD 56343         &2004-09-21&08:32:57 &$-$3609 &  69 &$-$3144 &  90 &$-$3415 &  54 & DDD &1950 &  M (new)  \\
BD-14 2015       &2005-01-29&05:47:41 & $-$31  &  72 &$-$290  & 258 & $-$32  &  69 & nnn &2400 &     \\
BD-14 2028       &2004-02-28&04:16:33 &   103  & 112 &$-$271  & 218 &    32  &  93 & nnn &1550 &     \\
                 &2004-03-08&04:45:24 &   119  & 130 &$-$394  & 240 &  $-$7  & 104 & nnn &1400 &     \\
HD 61045         &2003-02-08&04:28:15 &   412  &  63 &   476  & 108 &   506  & 111 & Ddd &2500 &  M (new)  \\
                 &2005-01-29&06:11:00 &   375  &  70 &   186  & 137 &   340  &  62 & DnD &2300 &  M  \\
BD-14 2040       &2003-02-08&04:28:15 &   169  & 273 &$-$149  & 244 &    16  & 166 & nnn & 750 &     \\
                 &2005-01-29&05:04:50 &    56  &  72 &$-$104  & 101 & $-$37  &  55 & nnn &2300 &     \\
HD 62376         &2005-01-29&02:41:28 &   101  &  62 &  $-$4  & 127 &    73  &  55 & nnn &2450 &     \\
CD-37 3845       &2003-02-08&04:52:12 &    98  & 113 &$-$430  & 201 &    10  &  97 & ndn &1450 &     \\
HD 62992         &2003-02-08&04:52:12 &$-$190  &  65 &$-$157  &  54 &$-$164  &  39 & ddd &2450 &  M (new)  \\
HD 63079         &2005-01-29&02:21:48 &     8  &  64 &   150  & 152 &    47  &  58 & nnn &2300 &     \\
HD 63401         &2003-02-08&00:36:57 &$-$532  &  61 &$-$551  & 119 &$-$527  &  53 & DDD &2900 & 6,M (new) \\
                 &2005-01-29&02:59:47 &   357  &  96 &   472  & 195 &  348   &  85 & Dnd &2450 &  M   \\
NGC 2489 58      &2005-01-30&06:10:58 &$-$325  & 205 &$-$826  & 313 &$-$449  & 170 & nnn &1200 &     \\
NGC 2489 40      &2005-01-30&06:10:58 &$-$489  & 220 &    36  & 251 &$-$387  & 158 & nnn & 850 &     \\
HD 65712         &2003-02-09&07:20:21 &$-$1025 &  84 &$-$1181 &  73 &$-$1111 &  54 & DDD &1800 &  M (new)  \\
                 &2005-01-29&04:22:56 &$-$668  &  55 &$-$357  &  74 &$-$558  &  43 & DDD &2450 &  M  \\
CPD-60944A       &2002-02-05&06:03:02 & $-$41  & 104 &   122  &  87 &    48  &  67 & nnn &1950 &7,R  \\
                 &2005-01-29&03:25:09 & $-$54  &  58 & $-$74  & 113 & $-$50  &  51 & nnn &2500 &7    \\
CPD-60944B       &2002-02-05&06:55:01 &   274  & 111 &   349  &  67 &   349  &  66 & nDD &1450 &7,R,m (new?) \\
                 &2005-01-29&03:52:30 &   143  &  63 &    93  &  89 &   158  &  50 & nnd &2650 &7    \\
CPD-60 978       &2002-02-04&06:08:09 &    48  & 103 & $-$35  &  71 &     2  &  58 & nnn &1750 & R   \\
                 &2002-02-05&04:19:29 &    58  &  83 &    39  &  71 &    46  &  54 & nnn &1850 & R   \\
                 &2004-03-08&05:25:21 & $-$91  & 103 &$-$229  & 189 & $-$83  &  88 & nnn &1650 &     \\
HD 65987         &2002-02-04&07:54:30 &   629  & 162 &   573  & 142 &   605  & 106 & DDD &1400 & R,M (new) \\
                 &2004-02-28&05:13:09 &$-$443  &  70 &$-$507  & 105 &$-$460  &  58 & DDD &2150 & M         \\
HD 66295         &2002-02-05&05:08:50 &$-$547  &  84 &$-$515  &  70 &$-$534  &  54 & DDD &2000 & R,M (new) \\
                 &2003-02-09&06:30:09 &   499  &  73 &   421  &  46 &   438  &  38 & DDD &2350 & R,M       \\
HD 66318         &2002-02-05&05:08:50 &  4410  & 105 &  5129  &  40 &  5044  &  37 & DDD &1600 &8,R,M (new)    \\
                 &2003-02-09&06:30:09 &  4266  &  85 &  5322  &  31 &  5196  &  29 & DDD &1850 & R,M  \\
NGC 2546 258     &2004-09-27&07:57:06 &$-$196  & 102 &$-$454  & 145 &$-$257  &  70 & ndD &1657 &     \\
                 &2005-01-30&07:12:58 & $-$83  &  98 &    16  &  78 & $-$21  &  55 & nnn &1850 &     \\
NGC 2546 201     &2005-01-29&06:52:31 &   307  &  62 &   355  &  57 &   594  & 158 & DDD &2650 &  M  (new) \\
CPD-37 1989      &2005-01-29&06:52:31 & $-$94  &  91 &    22  & 164 & $-$19  &  72 & nnn &2050 &     \\
HD 69004         &2004-09-27&08:53:02 &$-$100  &  54 &   110  & 116 & $-$58  &  48 & nnn &2900 &     \\
HD 69067         &2004-09-29&08:51:17 &   518  &  64 &   554  &  96 &   499  &  52 & DdD &2500 &  M  (new) \\
                 &2004-04-20&23:29:28 &   609  &  59 &   377  &  78 &   537  &  46 & DDD &2300 &  M  \\
HD 74169         &2003-02-08&05:13:58 &$-$151  &  57 &$-$191  &  56 &$-$175  &  39 & ndD &2250 &  m  (new?)\\
                 &2004-03-02&04:38:32 & $-$65  &  56 & $-$37  &  57 & $-$51  &  39 & nnn &2500 &     \\
HD 74535         &2003-02-08&05:33:45 & $-$83  &  68 &$-$126  & 137 & $-$96  &  60 & nnn &2250 &     \\
HD 75239         &2005-01-29&08:13:02 &   100  &  79 & $-$94  & 172 &    89  &  71 & nnn &2100 &     \\
HD 83002         &2005-01-30&07:45:58 &     5  &  80 &$-$102  & 161 & $-$18  &  70 & nnn &2000 &     \\
HD 87241         &2004-05-09&01:55:04 &    24  &  52 &$-$115  & 157 &    26  &  49 & nnn &2800 &     \\
HD 87240         &2003-02-08&07:15:26 &$-$257  &  58 &$-$239  & 103 &$-$259  &  50 & DnD &2500 &  M (new)  \\
HD 87266         &2003-02-08&06:21:28 & $-$45  &  38 &    44  &  54 &     0  &  32 & nnn &4100 &     \\
HD 304841        &2003-02-08&06:21:28 &$-$279  &  78 &$-$518  & 163 &$-$335  &  69 & DDD &1900 &  M (new)  \\
HD 304842        &2003-02-08&06:21:28 &    86  & 137 &$-$115  & 171 & $-$18  & 104 & nnn &2200 &     \\
HD 87405         &2003-02-08&06:21:28 & $-$82  &  45 &     2  &  65 & $-$61  &  36 & nnn &3850 &     \\
HD 89856         &2002-02-04&09:19:41 &$-$361  & 112 & $-$90  &  91 &$-$197  &  70 & Dnd &1650 &R,M (new)  \\
HD 92190         &2005-01-30&09:16:30 &    21  &  79 & $-$42  & 149 &    10  &  70 & nnn &2250 &     \\
HD 92385         &2005-01-29&09:28:11 &$-$580  &  60 &$-$266  & 142 &$-$519  &  55 & DnD &2550 &  M (new)  \\
Cr 228 30        &2004-05-28&00:02:29 &   692  & 143 &   405  & 212 &   558  & 113 & DnD &1550 &  M (new)  \\
HD 305451        &2004-05-28&00:02:29 &$-$181  &  89 & $-$96  & 258 &$-$149  &  87 & nnn &1950 &     \\
HD 96040         &2005-01-29&08:48:39 &$-$307  &  54 &$-$145  &  93 &$-$257  &  47 & DnD &2350 &  M (new)  \\
HD 96729         &2004-07-12&23:52:56 &  1125  &  71 &   601  & 100 &   949  &  57 & DDD &2100 &  M (new)  \\
HD 303821        &2005-01-30&08:35:12 & $-$74  &  88 &$-$113  &  84 &$-$100  &  60 & nnn &1750 &     \\
HD 122983        &2004-08-03&01:01:32 &   156  &  54 &   210  & 118 &   170  &  48 & dnD &3050 &     \\
HD 123183        &2004-07-13&00:32:44 &$-$440  & 146 &$-$181  & 473 &$-$409  & 176 & Dnn &1300 &  m (new?)  \\
HD 123225        &2004-07-13&00:32:44 &    97  &  83 &$-$176  & 200 & $-$22  &  72 & nnn &2400 &     \\
CPD-56 6330      &2003-02-08&08:22:37 &$-$108  & 125 &$-$404  & 165 &$-$191  &  99 & nnn &1350 &     \\
HD 127866        &2003-02-08&07:52:56 &$-$238  & 148 &$-$124  & 189 & $-$89  & 134 & nnn &1250 &     \\
CPD-57 7817      &2004-07-13&02:04:24 &$-$673  &  69 &$-$374  & 130 &$-$612  &  60 & DnD &2550 &  M (new)  \\
HD 146555        &2003-02-09&08:54:01 &   461  & 129 &   232  & 165 &   268  &  97 & dnn &1350 &  m (new?) \\
HD 149257        &2004-07-13&02:46:41 &   247  &  93 &    43  & 110 &   160  &  69 & nnn &2350 &     \\
HD 149277        &2004-07-13&02:46:41 &  2435  & 149 &  1549  & 186 &  2239  & 104 & DDD &2550 &  M (new)  \\
CD-48 11051      &2004-07-13&03:28:19 &$-$2604 & 231 &$-$1293 & 283 &$-$2013 & 117 & DDD &1400 &  M (new)  \\
                 &2004-07-24&02:51:35 &$-$2173 & 105 &$-$801  & 162 &$-$1772 &  74 & DDD &2050 &  M        \\
CD-48 11059      &2004-07-13&03:28:19 &   217  & 200 & $-$15  & 433 &   172  & 173 & nnn &1000 &     \\
                 &2004-07-24&02:51:35 &   193  & 138 &$-$145  & 183 &    90  & 107 & nnn &1400 &     \\
HD 322676        &2004-08-02&05:36:38 &$-$277  & 117 &   246  & 213 &$-$191  & 102 & nnn &2150 &     \\
HD 153948        &2004-08-02&05:36:38 &   199  &  58 &   176  & 141 &   195  &  53 & DnD &3100 &  M (new)  \\
NGC 6383 26      &2004-08-03&04:21:50 &$-$275  & 217 &   303  & 251 &    48  & 152 & nnn &1200 &     \\
HD 317857        &2004-07-24&03:52:22 &$-$1677 &  75 &$-$1393 &  81 &$-$1558 &  54 & DDD &2150 &  M (new)  \\
HD 318107        &2004-07-25&23:40:24 &  6519  &  55 &  3784  &  59 &  5216  &  39 & DDD &2600 & 9,M \\
HD 318100        &2004-08-17&01:20:22 &   390  &  71 &    93  & 151 &   345  &  64 & DnD &2100 &  M (new)  \\
CD-32 13119      &2004-08-03&02:21:41 & $-$29  &  65 & $-$39  &  37 & $-$37  &  30 & nnn &2450 &     \\
HD 318095        &2004-08-03&02:21:41 &   101  &  64 &$-$168  & 138 &    42  &  54 & nnn &2658 &     \\
HD 162305        &2004-08-03&06:18:16 &     1  &  40 &   122  &  95 &    24  &  37 & nnn &3950 &     \\
                 &2004-09-26&02:18:57 &   112  &  72 &   230  & 150 &   120  &  65 & nnn &2150 &     \\
HD 320764        &2004-08-23&01:44:37 & $-$89  &  78 & $-$43  & 154 & $-$69  &  66 & nnn &2400 &     \\
HD 162725        &2004-08-02&06:36:03 & $-$67  &  35 &    37  & 115 & $-$60  &  33 & nnn &4450 &     \\
HD 169959        &2004-07-06&03:48:34 &$-$541  &  58 &$-$302  & 128 &$-$486  &  52 & DnD &2550 &  M (new)  \\
HD 170054        &2004-08-27&02:43:53 &    16  &  72 &   361  & 153 &    64  &  68 & nnn &2300 &     \\
                 &2004-09-23&01:21:38 &   254  & 100 &$-$207  & 136 &    84  &  77 & nnn &2200 &     \\
BD-19 5044L      &2004-08-23&02:34:33 &$-$223  & 123 &$-$347  & 341 &$-$235  & 111 & nnn &1550 &     \\
BD-19 5046       &2004-08-23&02:34:33 &    68  & 100 &    64  &  64 &    60  &  92 & nnn &2650 &     \\
HD 170836        &2004-08-23&03:26:34 &$-$636  & 135 &$-$193  & 220 &$-$487  & 114 & DnD &1350 &  M (new)  \\
                 &2004-08-28&04:37:57 &$-$633  & 100 &$-$449  & 193 &$-$583  &  88 & DdD &1950 &  M  \\
                 &2004-09-26&02:54:12 &   502  &  95 &   436  & 140 &   483  &  77 & dDD &2100 &  M  \\
HD 170860A       &2004-08-04&06:40:14 & $-$70  &  79 &    62  & 152 & $-$41  &  69 & nnn &2050 &     \\
HD 190290        &2004-07-07&08:22:31 &  2420  & 110 &  2320  &  48 &  2341  &  41 & DDD &2000 &10,M \\
\hline
\end{longtable}
\vspace{0.2cm}

\noindent
R: observed with GRISM 600R\\
M: definite detection according to our judgement\\
(new): Star previously not known as magnetic\\
m: marginal detection according to our judgement\\
1: four null detections by Borra (1994; pr.\ comm. cited in Bychkov et al.\ \cite{Bycetal03}) \\
2: Sargent, Sargent, \& Strittmatter (\cite{Saretal67}) reported $\Bz=1300 \pm 400$\,G.
   Several additional measurements by Conti (\cite{Con70}) suggest that \Bz\ varies approximately
   from $-700$ to $+1400$\,G \\
3: Borra, Landstreet, \& Thompson (\cite{Boretal83}) measured $\Bz=628 \pm 178$\,G \\
4: Field detected by Sargent et al.\ (\cite{Saretal67}), Conti (\cite{Con70}), Borra et al.
   (\cite{Boretal83}), Mathys \& Hubrig (\cite{MatHub97}) \\
5: field detection reported by Bagnulo et al. (\cite{Bagetal04})\\
6: null detection by Bohelender, Landstreet, \& Thompson (\cite{Bohetal93}) \\
7: SIMBAD coordinates are inverted (A with B and B with A).
For CpD-60 944A RA \& DEC are 07:56:45.3  -60:48:55; 
for CpD-60 944B RA \& DEC are 07:56:46.3  -60:48:59.2\\
8: field detection reported by Bagnulo et al. (\cite{Bagetal03})\\
9: Mathys \& Hubrig (\cite{MatHub97}) detected $\Bz=1985 \pm 230$\,G \\
10: Hubrig et al. (\cite{Hubetal04}) detected $\Bz = 3220 \pm 73$\,G and
   $\Bz = 3250 \pm 111$\,G

%% file: 4223tableA4.tex
\begin{longtable}{lllr@{\,$\pm$}lr@{\,$\pm$}lr@{\,$\pm$}llrl}
\caption{\label{Table_Normals} \Bz\ measurements for stars not known to be
chemically peculiar stars. Unless otherwise stated in the notes, the
observations have been obtained with grism 600\,B.}  \\ 
\hline\hline
    &
    &     &\multicolumn{2}{c}{\Bz\ (G)}
          &\multicolumn{2}{c}{\Bz\ (G)}
          &\multicolumn{2}{c}{\Bz\ (G)}
          &    &     &      \\ 
\multicolumn{1}{c}{ID1} & 
\multicolumn{1}{c}{DATE}&
\multicolumn{1}{c}{UT\ \ \ \ \ \ } &
\multicolumn{2}{c}{Balmer lines} &
\multicolumn{2}{c}{metal lines}  &
\multicolumn{2}{c}{full spectrum}&
Flag& 
\multicolumn{1}{c}{SNR} & 
Note \\ 
\hline
\endfirsthead
\caption{continued.}\\
\hline\hline
    &
    &     &\multicolumn{2}{c}{\Bz\ (G)}
          &\multicolumn{2}{c}{\Bz\ (G)}
          &\multicolumn{2}{c}{\Bz\ (G)}
          &    &     &      \\ 
\multicolumn{1}{c}{ID1} & 
\multicolumn{1}{c}{DATE}&
\multicolumn{1}{c}{UT\ \ \ \ \ \ } &
\multicolumn{2}{c}{Balmer lines} &
\multicolumn{2}{c}{metal lines}  &
\multicolumn{2}{c}{full spectrum}&
Flag& 
\multicolumn{1}{c}{SNR} & 
Note \\ 
\hline
\endhead
\hline
\endfoot
HD 36559         &2003-02-08&01:40:07  &   $-$3  &    81 &  $-$76  &   146 &  $-$25  &    68 & nnn &1900&    \\
HD 36671         &2003-02-08&02:08:23  &     69  &   136 &     49  &    70 &     50  &    62 & nnn &1500&    \\
HD 37041         &2003-02-08&01:11:00  &    125  &    92 &     26  &    70 &     47  &    58 & nnn &3950&    \\
HD 37428         &2003-02-08&02:49:06  & $-$175  &    88 &  $-$45  &   158 & $-$159  &    88 & nnn &2100&    \\
HD 252214        &2003-02-08&03:40:54  &  $-$49  &    55 &  $-$11  &    68 &  $-$26  &    46 & nnn &3350&    \\
HD 41909         &2003-02-08&03:40:54  &      5  &    49 &    119  &    23 &    104  &    20 & nDD &2900&    \\
HD 252248        &2003-02-08&03:40:54  & $-$144  &   108 & $-$160  &   104 & $-$189  &    77 & nnn &2800&    \\
HD 252266        &2003-02-08&03:40:54  &    111  &    89 &  $-$94  &    78 &  $-$18  &    60 & nnn &2500&    \\
NGC 2244 336     &2003-02-09&03:52:60  &    399  &   182 &  $-$64  &    84 &  $-$19  &    74 & nnn &1250&    \\
CPD-20 1637      &2004-09-28&08:26:26  &     76  &    67 &  $-$26  &   194 &     88  &    62 & nnn &2700&    \\
BD-20 1571       &2004-09-28&08:26:26  & $-$104  &    81 &     58  &   129 &  $-$37  &    65 & nnn &2400&    \\
                 &2005-01-30&02:50:57  &    265  &   108 &     55  &   156 &    200  &    81 & dnd &1650&    \\
NGC 2287 AR 157  &2005-01-30&03:35:37  &  $-$14  &   138 & $-$153  &   216 &      2  &   106 & nnn &1250&    \\
CPD-20 1645      &2005-01-30&03:35:37  &  $-$55  &    88 &    130  &   171 &  $-$15  &    72 & nnn &1950&    \\
HD 52980         &2005-01-30&05:03:48  & $-$131  &    58 &  $-$48  &   173 & $-$111  &    54 & nnn &2450&    \\
BD-08 1708       &2005-01-30&05:03:48  &   $-$1  &   146 &  $-$65  &   335 &  $-$48  &   131 & nnn &1200&    \\
CSI-10-07049     &2002-02-05&01:22:04  & $-$234  &   157 &    161  &   134 &  $-$19  &    76 & nnn &1350&  R \\
BD-10 1875       &2002-02-05&01:22:04  &  $-$92  &   202 &   $-$8  &   188 &  $-$31  &   138 & nnn &1550&  R \\
NGC 2343 22      &2002-02-05&01:22:04  &    243  &   218 &     88  &   204 &    242  &   150 & nnn &1150&  R \\
NGC 2343 40      &2002-02-05&03:00:13  &    215  &   238 &     45  &   153 &     68  &   127 & nnn & 750&  R \\
CSI-10-07053 5   &2002-02-05&01:22:04  &    320  &   263 &    116  &   253 &    251  &   186 & nnn & 890&  R \\
BD-10 1878       &2002-02-05&03:00:13  &    111  &   115 & $-$144  &   145 &     12  &    89 & nnn &1450&  R \\
NGC 2343 25      &2002-02-05&03:00:13  &    280  &   246 &    124  &   224 &    291  &   192 & nnn &1000&  R \\
NGC 2343 16      &2002-02-05&03:00:13  & $-$225  &   183 & $-$162  &   186 & $-$227  &   127 & nnn &1350&  R \\
BD-10 1879a      &2002-02-04&03:50:42  & $-$178  &   228 &     12  &   171 &  $-$20  &   132 & nnn &1350&  R \\
NGC 2343 34      &2002-02-04&01:35:56  & $-$374  &   272 &    169  &   214 &  $-$40  &   167 & nnn & 800&  R \\
NGC 2343 43      &2002-02-04&05:01:44  &     64  &   334 &     34  &   243 &     10  &   181 & nnn & 700&  R \\
NGC 2343 28      &2002-02-05&03:00:13  &    164  &   234 &     24  &   181 &     82  &   145 & nnn &1050&  R \\
NGC 2343 37      &2002-02-04&05:01:44  &  $-$51  &   274 &     21  &   176 &  $-$59  &   146 & nnn & 850&  R \\
HD 54304         &2002-02-04&02:57:20  &    197  &   106 &  $-$47  &    99 &     54  &    70 & nnn &2000&  R \\
NGC 2343 27      &2002-02-04&05:01:44  &     40  &   262 & $-$103  &   212 &  $-$32  &   164 & nnn &1050&  R \\
NGC 2343 26      &2002-02-04&05:01:44  &     75  &   200 &  $-$11  &   203 &     32  &   140 & nnn &1200&  R \\
NGC 2343 6       &2002-02-04&02:57:20  & $-$132  &   107 &    113  &   121 &  $-$32  &    80 & nnn &1900&  R \\
NGC 2343 36      &2002-02-04&01:35:56  & $-$226  &   260 &    388  &   305 &     26  &   193 & nnn &1000&  R \\
NGC 2343 17      &2002-02-04&01:35:56  &     50  &   167 &  $-$11  &   212 &  $-$13  &   118 & nnn &1250&  R \\
NGC 2343 10      &2002-02-04&02:57:20  &     67  &   113 &    152  &   132 &    122  &    85 & nnn &1550&  R \\
NGC 2343 31      &2002-02-04&05:01:44  &  $-$71  &   280 & $-$530  &   284 & $-$296  &   192 & nnn & 750&  R \\
BD-10 1883B      &2002-02-04&02:57:20  &     17  &   215 &     70  &   185 &     55  &   139 & nnn &1250&  R \\
BD-10 1883A      &2002-02-04&01:35:56  &  $-$45  &   173 & $-$134  &   150 & $-$115  &   112 & nnn &1750&  R \\
NGC 2343 18      &2002-02-04&01:35:56  &     76  &   176 &    448  &   214 &    166  &   131 & nnn &1250&  R \\
NGC 2343 38      &2002-02-04&02:57:20  &    187  &   352 & $-$123  &   243 &  $-$23  &   198 & nnn & 700&  R \\
HD 54360         &2002-02-04&01:35:56  &  $-$74  &   166 &  $-$47  &   102 &  $-$50  &    85 & nnn &2150&  R \\
BD-10 1885B      &2002-02-04&05:01:44  &    340  &   136 & $-$102  &   115 &     89  &    88 & nnn &1750&  R \\
HD 54388         &2002-02-04&03:50:42  &      5  &    64 & $-$134  &    54 &  $-$66  &    41 & nnn &2450&    \\
HD 60940         &2005-01-29&05:47:41  &      6  &    50 &     97  &   165 &     10  &    47 & nnn &3250&    \\
HD 60996         &2005-01-29&05:47:41  &     61  &    58 &     93  &   157 &     76  &    57 & nnn &3250&    \\
HD 62974         &2003-02-08&04:52:12  &  $-$26  &   105 &   $-$3  &   205 &  $-$50  &    87 & nnn &1850&    \\
NGC 2489 59      &2005-01-30&06:10:58  &    360  &   169 &   1256  &   372 &    517  &   181 & ndn &1300&    \\
HD 65691         &2002-02-05&06:03:02  & $-$158  &   215 & $-$344  &   172 & $-$262  &   134 & nnn &1500&  R \\
CPD-60 942       &2002-02-05&06:03:02  &    616  &   281 & $-$208  &   216 &    106  &   169 & nnn & 950&  R \\
CD-60 1929       &2002-02-05&06:03:02  &  $-$52  &   109 &    244  &   108 &    110  &    76 & nnn &2100&  R \\
CD-60 1932       &2002-02-05&06:55:01  & $-$110  &   338 & $-$144  &   231 &  $-$99  &   188 & nnn &1100&  R \\
                 &2002-02-05&06:03:02  &  $-$28  &   305 &      7  &   225 &  $-$35  &   178 & nnn &1200&  R \\
HD 65869         &2002-02-05&06:55:01  &    233  &   118 & $-$203  &   107 &   $-$5  &    79 & nnn &2750&  R \\
HD 65896         &2002-02-04&07:54:30  &    109  &   111 &    144  &    98 &    127  &    73 & nnn &1900&  R \\
HD 65950         &2002-02-05&07:45:43  &  $-$32  &    50 & $-$123  &    62 & $-$114  &    50 & nnn &2800&  R \\
V373 Car         &2004-03-08&05:25:21  & $-$187  &    75 &  $-$34  &    61 & $-$104  &    48 & nnn &1350&    \\
                 &2002-02-05&04:19:29  &  $-$87  &    23 &  $-$25  &    36 &  $-$66  &    19 & DnD &1600&  R \\
HD 65949         &2002-02-04&07:54:30  &  $-$19  &    89 &    158  &    83 &     74  &    61 & nnn &2250&  R \\
BD-60 969        &2002-02-04&07:54:30  &     43  &   136 & $-$138  &   107 &  $-$69  &    84 & nnd &2350&  R \\
BD-60 975        &2002-02-05&04:19:29  &    145  &   111 &    100  &   111 &    127  &    78 & nnd &2250&  R \\
                 &2002-02-04&06:08:09  &     65  &   123 &    229  &   140 &    132  &    89 & nnn &2250&  R \\
CPD-60 977       &2002-02-04&07:01:42  &     35  &   109 &     75  &    94 &     57  &    70 & nnn &1550&  R \\
V410 Car         &2002-02-04&07:01:42  &     77  &   225 & $-$200  &   188 &  $-$98  &   140 & nnn &1450&  R \\
V392 Car         &2004-03-08&05:25:21  &    221  &   129 &  $-$75  &   144 &    111  &    91 & nnn &1250&    \\
                 &2002-02-05&04:19:29  &   $-$4  &    97 &     85  &    73 &     34  &    57 & nnn &2000&  R \\
                 &2002-02-05&07:45:43  & $-$285  &   220 &     82  &   142 &     15  &   119 & nnn &1200&  R \\
CD-60 1967       &2002-02-04&06:08:09  &     20  &    32 &     49  &    59 &     32  &    28 & nnn &2700&  R \\
CD-60 1971       &2002-02-04&06:08:09  &    252  &   149 &     46  &    96 &    101  &    80 & nnn &2600&  R \\
V417 Car         &2002-02-04&07:54:30  &  $-$78  &   130 & $-$175  &   112 &  $-$73  &    80 & nnn &1400&  R \\
CPD-60 984       &2002-02-04&07:01:42  &     33  &   137 &    170  &   103 &     52  &    77 & nnn &1500&  R \\
CPD-60 986       &2002-02-04&07:01:42  &  $-$62  &   160 &  $-$47  &   153 &  $-$38  &   109 & nnn &1825&  R \\
V418 Car         &2002-02-05&04:19:29  &   $-$9  &   222 &     98  &   172 &     68  &   133 & nnn &1190&  R \\
CD-60 1974       &2002-02-05&04:19:29  & $-$301  &   154 &  $-$24  &   119 & $-$126  &    94 & nnn &2100&  R \\
                 &2004-03-08&05:25:21  & $-$168  &   128 &    139  &   360 & $-$165  &   121 & dnd &1350&    \\
CD-60 1975       &2002-02-04&06:08:09  &      0  &   122 & $-$116  &   100 &  $-$48  &    76 & nnn &2100&  R \\
CD-60 1976       &2002-02-04&07:01:42  &  $-$50  &   182 &  $-$12  &   121 &  $-$58  &   104 & nnn &1850&  R \\
CD-60 1979       &2002-02-05&07:45:43  & $-$210  &   269 & $-$109  &   276 & $-$162  &   191 & nnn &1050&  R \\
CD-60 1978       &2002-02-04&07:01:42  &    145  &   216 &     96  &   244 &     67  &   161 & nnn &1300&  R \\
CD-60 1981       &2002-02-05&07:45:43  &    414  &   220 &  $-$77  &   161 &    167  &   141 & nnn &1000&  R \\
V420 Car         &2002-02-05&04:19:29  &  $-$29  &   215 & $-$227  &   172 & $-$127  &   134 & nnn &1500&  R \\
HD 66137         &2002-02-04&07:01:42  & $-$124  &   160 &    110  &   125 &     53  &   100 & nnn &1800&  R \\
HD 66194         &2002-02-04&06:08:09  & $-$183  &    97 & $-$237  &   353 & $-$197  &    93 & nnn & 600&  R \\
CD-60 1996       &2002-02-05&05:08:50  &    457  &   207 &    744  &   206 &    631  &   159 & nDD &1000&  R \\
                 &2003-02-09&06:30:09  &     81  &   169 &     98  &   147 &    108  &   120 & nnn &1200&    \\
CD-60 1999       &2003-02-09&06:30:09  &    100  &   289 &  $-$67  &    91 &  $-$53  &    87 & nnn &1050&    \\
                 &2002-02-05&05:08:50  &     80  &   233 & $-$289  &   205 &  $-$83  &   149 & nnn &1350&  R \\
$[$N$75]$ 195    &2005-01-30&07:12:58  &     68  &   103 &  $-$25  &    79 &     34  &    58 & nnn &2300&    \\
$[$N$75]$ 196    &2004-09-27&07:57:06  &    183  &   166 &    444  &   386 &    603  &   144 & nnd &1400&    \\
                 &2005-01-30&07:12:58  &  $-$40  &   162 & $-$170  &   202 &  $-$50  &   127 & nnn &1450&    \\
NGC 2546 272     &2004-09-27&07:57:06  &     26  &    96 &     61  &   207 &     22  &    86 & nnn &2350&    \\
                 &2005-01-30&07:12:58  &    188  &   126 &  $-$21  &   125 &     24  &    97 & nnn &2000&    \\
CD-37 4353       &2005-01-29&06:52:31  & $-$126  &    71 &  $-$30  &   346 & $-$102  &    69 & nnn &2450&    \\
HD 69003         &2004-09-27&08:53:02  &  $-$48  &    53 & $-$151  &   259 &  $-$42  &    51 & nnn &3500&    \\
HD 298537        &2005-01-30&07:45:58  & $-$221  &   334 & $-$402  &   238 & $-$293  &   178 & nnn & 750&    \\
HD 298536        &2005-01-30&07:45:58  &     49  &   196 & $-$105  &   508 &  $-$86  &   179 & nnn & 950&    \\
NGC 3114 233     &2003-02-08&07:15:26  & $-$123  &    96 &     53  &   191 &  $-$87  &    85 & nnn &1850&    \\
CPD-59 1698      &2003-02-08&06:21:28  &      1  &   126 & $-$333  &   406 &   $-$9  &   115 & nnn &1650&    \\
CPD-59 1700      &2003-02-08&07:15:26  &    287  &   161 &     32  &   240 &    212  &   133 & nnn &1400&    \\
CPD-59 1703      &2003-02-08&06:21:28  & $-$106  &   118 & $-$165  &   219 & $-$128  &    98 & nnn &1550&    \\
HD 298051        &2002-02-04&09:19:41  & $-$395  &   213 &  $-$61  &   245 & $-$250  &   160 & nnn &1150&  R \\
HD 89901         &2002-02-05&08:36:32  & $-$153  &   163 &    250  &   156 &     39  &   113 & nnn &2050&  R \\
HD 89900         &2002-02-05&08:36:32  &   $-$3  &   136 &   $-$1  &   135 &      0  &    96 & nnn &2050&  R \\
HD 89915         &2002-02-05&08:36:32  &     37  &   176 & $-$189  &   121 & $-$113  &   100 & nnn &2600&  R \\
HD 89922         &2002-02-04&09:19:41  &  $-$47  &   130 &  $-$23  &   112 &  $-$46  &    84 & nnn &1800&  R \\
HD 298047        &2002-02-05&08:36:32  &      1  &   179 &    154  &   233 &     94  &   143 & nnn &1550&  R \\
HD 89938         &2002-02-05&08:36:32  & $-$152  &   181 &  $-$10  &   206 &  $-$91  &   136 & nnn &1600&  R \\
HD 89937         &2002-02-05&08:36:32  &     14  &   190 &    120  &   154 &     71  &   120 & nnn &1600&  R \\
HD 89956         &2002-02-05&09:20:10  & $-$374  &   337 &     20  &   222 &  $-$81  &   185 & nnn &1350&  R \\
CPD-51 3249      &2003-02-09&08:13:16  &  $-$54  &   200 &  $-$16  &   173 &  $-$51  &   138 & nnn &1300&    \\
HD 298053        &2003-02-09&08:13:16  &     60  &    98 &    105  &   111 &    116  &    84 & nnn &1700&    \\
HD 303107        &2005-01-30&09:16:30  & $-$253  &   143 &    915  &   430 & $-$190  &   133 & nnn &1350&    \\
HD 305535        &2004-05-28&00:02:29  &  $-$26  &    64 &    225  &   123 &     30  &    55 & nnn &3000&    \\
HD 305543        &2004-05-28&00:02:29  &   $-$4  &   224 & $-$149  &   121 & $-$105  &    90 & nnn &2700&    \\
HD 96653         &2004-07-12&23:52:56  &     87  &    45 &    824  &   486 &     96  &    44 & nnn &1600&    \\
NGC 3532 447     &2004-07-12&23:52:56  &  $-$70  &   101 &    456  &   324 &   $-$3  &    95 & nnn &2000&    \\
HD 96790         &2004-07-12&23:52:56  & $-$168  &   174 & $-$118  &   126 & $-$159  &    98 & nnn &2150&    \\
CD-47 8868       &2004-08-03&01:01:32  &      3  &    89 & $-$284  &   290 &  $-$16  &    84 & nnn &2000&    \\
HD 123201B       &2004-07-13&00:32:44  & $-$109  &    96 &    353  &   281 &  $-$77  &    94 & nnn &1700&    \\
HD 127835        &2003-02-08&09:03:56  &     87  &    85 &   $-$5  &   184 &     77  &    78 & nnn &2050&    \\
CPD-56 6334      &2003-02-08&08:22:37  &     80  &   140 & $-$191  &   158 &  $-$33  &    92 & nnn &1950&    \\
NGC 5662 126     &2003-02-08&08:22:37  &    233  &   141 &    104  &   172 &    187  &   109 & nnn &1450&    \\
HD 127900        &2003-02-08&09:03:56  &   $-$7  &    56 &    236  &   126 &     30  &    52 & nnn &2700&    \\
HD 127924        &2003-02-08&09:03:56  &     20  &   100 &  $-$22  &   125 &     25  &    88 & nnn &2050&    \\
NGC 6087 129     &2004-07-13&02:04:24  &   $-$5  &    69 &    133  &   175 &     46  &    62 & nnn &2750&    \\
TYC 8719- 717-1  &2004-07-13&02:04:24  &      0  &    84 &    266  &   171 &     51  &    74 & ndn &2300&    \\
HD 146484        &2003-02-09&08:54:01  & $-$247  &   100 & $-$316  &   258 & $-$222  &    98 & nnd &1600&    \\
CD-48 11050      &2004-07-13&03:28:19  & $-$148  &   114 &    129  &   325 &  $-$47  &    98 & nnn &1550&    \\
                 &2004-07-24&02:51:35  &  $-$55  &    74 &     26  &   187 &  $-$11  &    63 & nnn &2200&    \\
CD-48 11060      &2004-07-13&03:28:19  & $-$143  &   172 &     59  &   310 & $-$102  &   136 & nnn &1250&    \\
                 &2004-07-24&02:51:35  &  $-$99  &   112 & $-$228  &   161 &  $-$95  &    87 & nnn &1850&    \\
HD 323673        &2004-08-02&05:36:38  &  $-$43  &    91 &     97  &   210 &      1  &    81 & nnn &2150&    \\
NGC 6383 28      &2004-08-03&04:21:50  &    134  &   145 &  $-$30  &   448 &     17  &   143 & nnn &1250&    \\
NGC 6383 700     &2004-08-03&04:21:50  &    177  &   186 &     62  &   187 &    148  &   116 & nnn &1250&    \\
NGC 6383 87      &2004-08-03&04:21:50  &     78  &   137 &     68  &   175 &     35  &   106 & nnn &1250&    \\
HD 317846        &2004-07-24&03:52:22  &  $-$20  &    74 &  $-$77  &   193 &  $-$23  &    68 & nnn &2700&    \\
NGC 6383 102     &2004-07-24&03:52:22  &    163  &   109 & $-$221  &   327 &    147  &   101 & nnn &1750&    \\
HD 317852        &2004-08-03&04:21:50  &  $-$61  &   131 &     34  &    71 &  $-$28  &    65 & nnn &1700&    \\
HD 318108        &2004-07-25&23:40:24  & $-$131  &    79 & $-$156  &   350 &  $-$98  &    82 & nnn &2150&    \\
HD 318109        &2004-07-25&23:40:24  &     67  &    87 &     62  &   442 &     93  &    90 & nnn &1950&    \\
CD-32 13089      &2004-08-17&01:20:22  &  $-$19  &    94 &     76  &   126 &     57  &    68 & nnn &2000&    \\
CD-32 13093      &2004-08-17&01:20:22  & $-$156  &   113 & $-$406  &   344 & $-$245  &   106 & nnn &1700&    \\
V976 Sco         &2004-08-17&01:20:22  &  $-$62  &   214 & $-$199  &   279 &  $-$84  &   142 & nnn & 950&    \\
HD 318099        &2004-08-17&01:20:22  &  $-$40  &    84 &    587  &   363 &  $-$38  &    89 & nnn &1850&    \\
HD 320765        &2004-08-23&01:44:37  &  $-$18  &    71 &    281  &   235 &      0  &    65 & nnn &2500&    \\
HD 162678        &2004-08-02&06:36:03  &   $-$2  &    36 &  $-$14  &   132 &      3  &    34 & nnn &4150&    \\
HD 162724        &2004-08-02&06:36:03  &  $-$33  &    54 & $-$329  &   171 &  $-$53  &    51 & nnn &5200&    \\
BD-19 5044 F     &2004-08-23&02:34:33  &     94  &   100 &   $-$2  &   255 &     63  &    96 & nnn &2050&    \\
BD-19 5045 ?     &2004-08-23&02:34:33  &  $-$12  &    74 &    191  &   120 &     86  &    60 & nnn &2650&    \\
BD-19 5044 M     &2004-08-23&02:34:33  &  $-$50  &   127 &    244  &   492 &  $-$32  &   120 & nnn &1450&    \\
HD 170835        &2004-08-23&03:26:34  & $-$242  &   239 & $-$500  &   310 & $-$417  &   183 & nnn &1500&    \\
                 &2004-08-28&04:37:57  &  $-$15  &   200 & $-$635  &   235 & $-$435  &   142 & nnD &2100&    \\
\hline
\end{longtable}

%% file: 4223tableA5.tex
\begin{longtable}{l@{\ \ }l@{\ \ }r@{\ \ }c@{\ \ }l@{\ \ }lr@{\,$\pm$}llrl}
\caption{\label{Table_NonA} \Bz\ measurements for stars not of early spectral type.}\\
\hline\hline
    &     &    &Spectral&
    &     &\multicolumn{2}{c}{\Bz\ (G)}
          &    &     &      \\ 
\multicolumn{1}{c}{ID1} & 
\multicolumn{1}{c}{ID2} &
\multicolumn{1}{c}{$V$} &
\multicolumn{1}{c}{type}&
\multicolumn{1}{c}{DATE}&
\multicolumn{1}{c}{UT\ \ \ \ \ \ } &
\multicolumn{2}{c}{full spectrum}  &
Flag& 
\multicolumn{1}{c}{SNR} & 
Note \\ 
\hline
\endfirsthead
\caption{continued.}\\
\hline\hline
    &     &    &Spectral&
    &     &\multicolumn{2}{c}{\Bz\ (G)} 
          &    &     &      \\ 
\multicolumn{1}{c}{ID1} & 
\multicolumn{1}{c}{ID2} &
\multicolumn{1}{c}{$V$} &
\multicolumn{1}{c}{type}&
\multicolumn{1}{c}{DATE}&
\multicolumn{1}{c}{UT\ \ \ \ \ \ } &
\multicolumn{2}{c}{full spectrum}&
Flag& 
\multicolumn{1}{c}{SNR} & 
Note \\ 
\hline
\endhead
\hline
\endfoot
NGC 2244 330     &NGC 2244 PS 537   &10.8  &  K5    &2003-02-09&03:52:60  &     90  &    18 &d &2300&   \\
NGC 2244 365     &GSC 00154-00486   &11.9  &  G2    &2003-02-09&03:52:60  &      0  &    51 &n &1600&   \\
NGC 2244 364     &GSC 00154-00074   &12.2  &  --    &2003-02-09&03:52:60  &     53  &    35 &n &1200&   \\
NGC 2287 AR 155  &--                & --   &  --    &2005-01-30&03:35:37  &  $-$32  &   103 &n &1300&   \\
NGC 2343 15      &CSI-10-07056 2    & 11.4 &   [K4?]&2002-02-04&01:35:56  &     28  &    35 &n &1550& R \\
NGC 2343 23      &CSI-10-07057 5    & 11.9 &  [F2?V]&2002-02-04&02:57:20  &  $-$55  &    66 &n &1050& R \\
NGC 2343 19      &CSI-10-07058 2    & 11.5 &  [G/K?]&2002-02-04&01:35:56  & $-$220  &   121 &n &1350& R \\
HD 54387         &BD-10 1885        &  8.5 &  G5    &2002-02-04&01:35:56  &     69  &    26 &n &1500& R \\
HD 60968         &BD-14 2017        & 9.2  &  K0    &2005-01-29&05:47:41  &  $-$47  &    25 &n &2600&   \\
NGC 2516 DAC 515 &NGC 2516 SBL 481  &11.3  &  --    &2002-02-05&07:45:43  &    167  &   166 &n & 750& R \\
CD-60 1997       &NGC 2516 SBL 953  &10.4  &[F2/3?V]&2002-02-05&05:08:50  &     74  &    48 &n &1150& R \\
                 &                  &      &        &2003-02-09&06:30:09  &    107  &    37 &d &1450&   \\
HD 298045        &CD-51 4704        & 9.4  &  M3    &2003-02-09&08:13:16  &    190  &    32 &D &1750&   \\
HD 298054        &CD-51 4707        & 9.8  &  G0    &2003-02-09&08:13:16  &     29  &    16 &n &1750&   \\
CPD-58 3151      &NGC 3532 198      &10.0  &  [G0/2]&2004-07-12&23:52:56  &  $-$27  &    59 &n &1800&   \\
CD-47 8861       &NGC 5460 38       & 9.8  &  --    &2004-08-03&01:01:32  &  $-$59  &    34 &n &2100&   \\
NGC 5460 73      &CD-47 8891        & 9.6  &  K5    &2004-07-13&00:32:44  &     55  &    48 &n &1400&   \\
HD 127753        &CD-55 5722        & 7.1  &  K5III &2003-02-08&07:52:56  &     44  &    15 &n &1650&   \\
NGC 5662 118     &CSI-56-14321 1    &10.7  &  [K5V] &2003-02-08&08:22:37  &  $-$22  &    35 &n &1200&   \\
NGC 5662 CLB 149 &GSC 08687-01270   &11.5  &    [?] &2003-02-08&09:03:56  &    153  &    61 &n & 550&   \\
CPD-57 7883      &PPM 772347        & 9.6  &  --    &2003-02-09&08:54:01  &     69  &    87 &n &1100&   \\
CD-34 11864      &NGC 6396 PPM 78   &10.6  &  M4    &2003-02-09&09:27:31  &     96  &    39 &n & 700&   \\
NGC 6405 322     &--                & --   &  --    &2004-08-03&02:21:41  &     64  &    49 &n &2000&   \\
\hline
\end{longtable}